\documentclass[11pt,twoside]{article}
\usepackage{atmp}
\copyrightnotice{2002}{6}{107}{139}

\newtheorem{theorem}{Theorem}[section]

\newtheorem{lemma}[theorem]{Lemma}
\newtheorem{proposition}[theorem]{Proposition}

\numberwithin{equation}{section}

\newcommand{\R}{\mathbb{R}}
\newcommand{\Complex}{\mathbb{C}}
\renewcommand{\Re}{\mathop{\mathrm{Re}}}
\renewcommand{\Im}{\mathop{\mathrm{Im}}}

\newcommand{\longto}{\longrightarrow}
\newcommand{\norm}[1]{\left\Vert #1 \right\Vert}
\newcommand{\abs}[1]{\left| #1 \right|}
\newcommand{\bka}[1]{\left \langle #1 \right \rangle}
\newcommand{\bkA}[1]{\left \langle #1 \right \rangle}
\newcommand{\bke}[1]{\left( #1 \right)}
\newcommand{\bkt}[1]{\left[ #1 \right]}
\newcommand{\bket}[1]{\left\{ #1 \right\}}

\newcommand{\myproof}{\textsc{Proof:}\ }
\newcommand{\myendproof}{\hspace*{\fill}
  {{\bf \small Q.E.D.}} \vspace{10pt}}

\newcommand{\Pc}{\, \mathbf{P}\! _\mathrm{c} \, \! }

\newcommand{\PcH}{\, \mathbf{P}\! _\mathrm{c} \! ^{H_0} \,}

\newcommand{\Hc}{\, \mathbf{H} _\mathrm{c} \, \! }

\newcommand{\e}{\varepsilon}
\newcommand{\loc}{_{\mathrm{loc}}}
\newcommand{\al}{\alpha}

\newcommand{\la}{\lambda}

\newcommand{\wt}{\widetilde}

\newcommand{\pd}{\partial}
\newcommand{\donothing}[1]{}

\newcommand{\les}{\lesssim}
\newcommand{\lbar}[1]{\underline{#1}}

\newcommand{\dt}{\Delta t}
\newcommand{\nxi}{\alpha}

\begin{document}

\title{Classification of Asymptotic Profiles for
Nonlinear Schr\"odinger Equations with Small Initial Data}

\url{abs/math-ph/0205015}
\vspace*{-.3in}

\author{Tai-Peng Tsai}
\vspace*{-.5in}

\address{Institute for Advanced Study, Princeton, NJ 08540}
\addressemail{ttsai@ias.edu}

\author{Horng-Tzer Yau}
\vspace*{-.5in}

\address{Courant Institute, New York University, New York, NY 10012}
\addressemail{yau@cims.nyu.edu}

\markboth{\it CLASSIFICATION OF ASYMPTOTIC PROFILES FOR NLS}
{TAI-PENG TSAI and HORNG-TZER YAU}



\begin{abstract}

We consider a nonlinear Schr\"odinger equation  with a bounded
local potential in $\R^3$. The linear Hamiltonian is assumed to
have two bound states with the eigenvalues satisfying some
resonance condition.  Suppose that the initial data are localized
and small  in $H^1$. We prove that exactly three local-in-space
behaviors can occur as the time tends to infinity: 1. The
solutions vanish; 2. The solutions converge to nonlinear ground
states; 3.  The solutions converge to nonlinear excited states.
We also obtain upper bounds for the relaxation in all three cases.
In addition, a matching lower bound  for the relaxation to nonlinear
ground states
was given for a large set of initial data which is believed to be generic.
Our proof is based on outgoing estimates of the dispersive waves
which measure the relevant time-direction dependent
information of the dispersive wave. These estimates,
introduced in \cite{TY2}, provides the first  general notion
to measure the out-going tendency of waves in the setting of nonlinear
Schr\"odinger equations.

\end{abstract}

\cutpage

\section{Introduction}

Consider the nonlinear  Schr\"odinger equation
\begin{equation} \label{Sch}
i \pd _t \psi = (-\Delta + V) \psi + \la |\psi|^2 \psi, \qquad
\psi(t=0)= \psi_0 ,
\end{equation}
where $V$ is a smooth localized real potential,  $\la=\pm 1 $ and
$\psi=\psi(t,x):\R\times \R^3 \longto \Complex$ is a wave
function. For any solution $\psi(t)\in H^1(\R^3)$ the $L^2$-norm
and the Hamiltonian
\begin{equation} \label{1-2}
{\cal H}[\psi] = \int \frac 12 |\nabla \psi|^2 +  \frac 12 V
|\psi|^2 + \frac 14 \la |\psi|^4 \, d x
\end{equation}
are constants for all $t$. The global well-posedness for small
solutions in $H^1(\R^3)$ can be proved using these conserved
quantities and a continuity argument.
We assume that the linear Hamiltonian $H_0 :=- \Delta + V$ has two
simple eigenvalues $e_0<e_1<0$ with normalized eigen-functions
$\phi_0$, $\phi_1$. The nonlinear bound states to the
Schr\"odinger equation \eqref{Sch} are solutions to the equation
\begin{equation}   \label{Q.eq}
    (-\Delta + V) Q + \la |Q|^2 Q = EQ   .
\end{equation}
They are critical points to the Hamiltonian ${\cal H}[\psi] $
defined in \eqref{1-2} subject to the constraint that the
$L^2$-norm of $\psi$ is fixed. For any nonlinear bound state $Q=Q_E$,
$\psi(t) = Q e^{-i E t } $ is a solution to the nonlinear
Schr\"odinger equation.

We may obtain two families of such bound states by standard
bifurcation theory, corresponding to the two eigenvalues of  the
linear Hamiltonian. For any $E$ sufficiently close to $e_0$ so
that $E-e_0$ and $\la$ have the same sign, there is a unique
positive solution $Q=Q_E$ to \eqref{Q.eq} which decays
exponentially  as $x \to \infty$. See Lemma 2.1 of \cite{TY2}. We
call this family the {\it nonlinear ground states} and we refer to
it as $\bket{Q_{E}}_{E}$. Similarly, there is a {\it nonlinear
excited state} family $\bket{Q_{1,E_1}}_{E_1}$ for $E_1$ near
$e_1$.  We will abbreviate
them as $Q$ and $Q_{1}$. From the same Lemma 2.1 of \cite{TY2},
these solutions are small, localized and $\norm{Q_{E}} \sim
|E-e_0|^{1/2}$ and $\norm{Q_{1,E_1}} \sim |E_1-e_1|^{1/2}$.

\bigskip
Our goal is to classify the asymptotic dynamics for small initial data.
We have  proved \cite{TY3}
that there exists a family of  ``finite co-dimensional
manifolds" in the space of initial data
so that the dynamics asymptotically converge to
some excited states.
Outside a small
neighborhood of these manifolds, the asymptotic profiles
are given by some ground states \cite{TY2}.
In this article,
we shall extend the result of \cite{TY2} and prove that
the possible asymptotic profiles are either vacuum (i.e., vanishing
in $L^\infty$ norm), the ground states or the excited states.
Furthermore, we obtain the rates of the convergence for all cases.

\bigskip

We first state the assumptions on the potential $V$,
which is the same as in \cite{TY}.
Denote by  $L^2_{r}$ the weighted $L^2$ spaces ($r$ may be
positive or negative)
\begin{equation} \label{L2r.def}
L^2_{r}\,(\R^3) \; \equiv \; \bket{\phi \in L^2(\R^3) \; : \;
\bkA{x}^r \phi \in L^2(\R^3) }  .
\end{equation}
The space for initial data in \cite{TY} is
\begin{equation} \label{Y.def}
Y \equiv H^1(\R^3) \cap L^2_{r_0}\,(\R^3)  ,
 \qquad r_0>3  .
\end{equation}
We shall use $L^2\loc$ to denote $L^2 _{- r_0}$.  The parameter
$r_0>3$ is fixed and we can choose, say, $r_0=4$ for the rest of
this paper.

\noindent {\bf Assumption A0}: $- \Delta + V$ acting on
$L^2(\R^3)$ has $2$ simple eigenvalues $e_0<e_1< 0 $, with
normalized eigenvectors $\phi_0$ and $\phi_1$.

\noindent {\bf Assumption A1}: Resonance condition. Let $e_{01}=
e_1-e_0$ be the spectral gap of the ground state. We assume that
$2e_{01} > |e_0|$, i.e., $e_0<2 e_1$. Let
\begin{equation}\label{gamma0.def}
\gamma_0 := \lim_{\sigma \to 0+} \Im \bke{\phi_0 \phi_1^2, \,
\frac 1 {H_0 + e_0 - 2 e_1 -\sigma i} \Pc^{H_0}\phi_0 \phi_1^2}.
\end{equation}
Since the expression is quadratic, we have $\gamma_0 \ge 0$. We
assume, for some $s_0>0$,
\begin{equation}\label{A:gamma0}
\inf_{|s|<s_0} \lim_{\sigma \to 0+} \Im \bke{\phi_0 \phi_1^2, \,
\frac 1 {H_0 + e_0 - 2 e_1 +s -\sigma i} \Pc^{H_0}\phi_0 \phi_1^2}
\ge \frac 34 \gamma_0 > 0  .
\end{equation}
We shall use $0i$ to
replace $\sigma i$ and the limit $\lim_{\sigma \to 0+}$ later on.

\noindent {\bf Assumption A2}: For $\la Q_E^2$ sufficiently small,
the bottom of the continuous spectrum to $-\Delta+ V+\la Q_E^2$,
$0$, is not a generalized eigenvalue, i.e., not a resonance. Also,
we assume that $V$ satisfies the assumption in  \cite{Y} so that
the $W^{k,p}$ estimates  $k\le 2$ for the wave operator
$W_{H_0}=\lim_{t\to \infty} e^{i t H_0}e^{i t\Delta}$ hold for
$k \le 2$, i.e., there is  a small $\sigma>0$ such that,
\[
|\nabla^\beta V(x)| \le C \bkA{x}^{-5-\sigma}, \qquad \text{for }
|\beta|\le 2  .
\]
Also, the functions $(x\cdot \nabla)^k V$, for $k=0,1,2,3$, are
$-\Delta$ bounded with a $-\Delta$-bound $<1$:
\begin{equation*}
  \norm{(x\cdot \nabla)^k V\phi}_2 \le \sigma_0 \norm{-\Delta\phi}_2 +
C\norm{\phi}_2,
  \qquad \sigma_0 < 1 , \quad k=0,1,2,3   .
\end{equation*}

%
%
The main assumption in A0-A2
is the condition $2e_{01} > |e_0|$ in A1.
It guarantees that twice the excited state energy of $H_0 - e_0$
becomes a resonance in the continuum spectrum (of $H_0 - e_0$).
This resonance produces the main relaxation mechanism. If this
condition fails, the resonance occurs in higher order terms and a
proof of relaxation will be much more complicated. Also, the rate
of decay will be different.
\bigskip

%

%

\medskip

\begin{theorem}\label{th:1-2}
There is a small number $n_0>0$ such that if $ \|\psi_0\|_Y :=
\alpha \le n_0$ then exactly three possible long time dynamics may
occur as $t \to \infty$:

I. $\norm{\psi(t)} _{L^2\loc}  \le C_{\psi_0} t^{-6/11}$;

II. $\norm{\psi(t) - Q_{E}e^{-i E t + i \omega(t)}}_{L^2\loc} \le
C_{\psi_0} t^{-1/2} $ for some nonlinear ground state $Q_{E}\not =
0$ and real function $\omega(t)=O(\log t)$;

III. $\norm{\psi(t) - Q_{1,E_1}e^{-i E_1 t + i
\omega(t)}}_{L^2\loc} \le C_{\psi_0} t^{-1/2}$ for some nonlinear
excited state $Q_{1, E_1}\not = 0$ and real function
$\omega(t)=O(\sqrt t)$.

\end{theorem}

\medskip

Sufficient conditions guaranteeing  the convergence to the vacuum
(type I), the ground states (type II), or the excited states (type
III) are provided in \cite{TY2, TY3}.
The  type I or  III solutions constructed in \cite{TY3} are finite
co-dimensional subset of all small solutions, i.e.,  the initial
data for these solutions form a finite co-dimensional subset of
$\{\psi_0: \|\psi_0\|_Y \le n_0 \}$, the set of all small  initial
data. We believe that the type I or  III solutions in general can
be constructed in this way and thus have measure zero. The decay
rates of the type I or  III solutions constructed in \cite{TY3}
are of order $t^{-3/2}$; the corresponding upper bounds provided
in Theorem \ref{th:1-2} are $t^{-6/11}$ or $t^{-1/2}$. Since we
believe that all type I or III solutions  originate from the
construction in \cite{TY3}, these bounds are far from optimal.
They result from technical considerations of our classification
scheme (which will be explained in the following).

{The upper bound $t^{-1/2}$ obtained for the type II solutions in
Theorem~\ref{th:1-2} are optimal for initial data considered in
\cite{TY2},  where an lower bound of the same order was provided}.
However, there exists a measure zero set \cite{TY} such that the
decay rate is at least of order $t^{-3/2}$. We believe that all
solutions decaying to the ground states faster than $t^{-1/2}$
have measure zero. Summarizing,  we believe that the type II
solutions with decay rate exactly of order $t^{-1/2}$ are generic;
all other behavior are of measure zero.

There is a vast literature concerning the classification of
asymptotic dynamics for nonlinear Schr\"odinger equations with
small initial data. We shall only be able to mention a few: the
one bound state case \cite{BP1, C1, PW, SW1}, the one
dimension with two bound states \cite{BP, BS},  the three
dimension  with two bound states \cite{C2}  and \cite{SW3} where
results similar to Theorem \ref{th:1-2} were considered. Earlier
works concerning the related linear analysis were obtained in
\cite{G0,G, SS, SS2}.

To explain the idea for the proof, we decompose the wave
function using the eigenspaces of  the Hamiltonian  $H_0$ as
\begin{equation} \label{psidec0} \psi = \lbar{x}\phi_0 +
\lbar{y}\phi_1 +\lbar{\xi} , \quad \lbar{\xi} = \Pc^{H_0}\, \psi.
\end{equation}
This decomposition is not suitable for estimation and will be replaced by the
decomposition \eqref{psidec1} emphasizing the role of the excited states
in the next section. It is useful for the following heuristic explanation.

The key ingredients  for proving Theorem \ref{th:1-2} were
originated from the previous work \cite{TY, TY2}. Apart from the standard arguments
based on the normal form and resonance decay, the main new idea
introduced in \cite{TY2} was the concept of  outgoing estimates.
This concept allows us to capture the time-direction
dependent information of waves. Therefore, even though
the $L^2$ norm
of the dispersive wave may not change much in the time evolution,
its ``size" will decay in time when measured in terms of ``outgoing estimates"
(see Propositions \ref{th:t1}, \ref{th:t2}, \ref{th:t4}). Thus after certain
initial time,
the wave function will fall essentially into the region
considered in \cite{TY2} provided that it does not converge
to some excited state or the vacuum.
Hence we set up the following  flow chart:
\bigskip

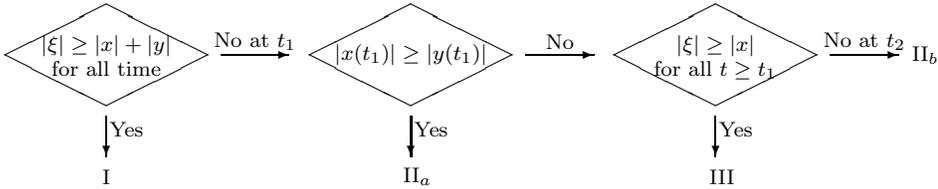
\setlength{\unitlength}{0.9mm}\noindent
\begin{picture}(105,30)

\put (0,20){\line(2,1){15}}
\put (0,20){\line(2,-1){15}}
\put (30,20){\line(-2,1){15}}
\put (30,20){\line(-2,-1){15}}

\put (32,20){\vector(1,0){10}}
\put (15,11){\vector(0,-1){6}}
\put(15,2){\makebox(0,0)[c]{\scriptsize I}}

\put(15,19.5){\makebox(0,0)[c]{\scriptsize
$\begin{array}{c} |\xi| \ge |x|+|y|\\  \text{for all time}\end{array}$}}

\put(18,9){\makebox(0,0)[c]{\scriptsize Yes} }
\put(37,22){\makebox(0,0)[c]{\scriptsize No at $t_1$} }

\put (45,20){\line(2,1){15}}
\put (45,20){\line(2,-1){15}}
\put (75,20){\line(-2,1){15}}
\put (75,20){\line(-2,-1){15}}

\put (77,20){\vector(1,0){10}}
\put (60,11){\vector(0,-1){6}}
\put(61,2){\makebox(0,0)[c]{\scriptsize II$_a$}}
\put(60,20){\makebox(0,0)[c]{\scriptsize $|x(t_1)|\ge |y(t_1)|$}}
\put(63,9){\makebox(0,0)[c]{\scriptsize Yes} }
\put(82,22){\makebox(0,0)[c]{\scriptsize No} }

\put (90,20){\line(2,1){15}}
\put (90,20){\line(2,-1){15}}
\put (120,20){\line(-2,1){15}}
\put (120,20){\line(-2,-1){15}}

\put (122,20){\vector(1,0){10}}
\put (105,11){\vector(0,-1){6}}
\put(106,2){\makebox(0,0)[c]{\scriptsize III}}


\put(105,19.5){\makebox(0,0)[c]{\scriptsize
$\begin{array}{c} |\xi| \ge |x| \\   \text{for all } t\ge t_1\end{array}$}}

\put(108,9){\makebox(0,0)[c]{\scriptsize Yes} }
\put(127,22){\makebox(0,0)[c]{\scriptsize No at $t_2$} }

\put(136,20){\makebox(0,0)[c]{\scriptsize II$_b$}}

\end{picture}\donothing{
{\small\begin{align*}
& \text{I: Dispersion dominated region. Convergence to the vacuum.} \\
& \text{II$_a$, II$_b$: Nonlinear ground states dominated region.
Convergence to nonlinear ground states.}\\
& \text{III:
Nonlinear excited states dominated region.
Convergence to nonlinear excited states.}
\end{align*}}
}
{\scriptsize
\begin{enumerate}
\item[I:] Dispersion dominated region. Convergence to the vacuum.

\item[II$_a$, II$_b$:] Nonlinear ground states dominated region.
Convergence to nonlinear ground states.

\item[III:]
Nonlinear excited states dominated region.
Convergence to nonlinear excited states.
\end{enumerate}
}

\bigskip

We first ask the question whether the dispersive part $\xi$ dominates
for all time. If it is, the dynamics will  converge to the vacuum,
the case I. If the dynamic fail this test at $t_1$, we then ask the second question
whether $|x(t_1)|\ge |y(t_1)|$. If yes, the ground state component dominates
and we are in the region II$_a$, which was
considered in \cite{TY} (formulated in a stronger form in \cite{TY2}).
Otherwise, the excited component dominates. We then test again whether
the dispersive wave dominates the
ground state component for all time $t\ge t_1$.
If yes, this produces the excited state dominated region III.
Otherwise, we reach the region II$_b$ at the time $t_2$.  At this point both
the ground state component
and the $L^\infty$
norm of the dispersive wave can be arbitrarily small
compared with the excited state
component. Furthermore, the $L^2$ norm of the dispersive wave can be
much larger than even the excited state component.
In other words, we may have
$$
\|\xi\|_{L^2} \gg |y| \gg |x| \gg \|\xi\|_{L^\infty}
$$
Notice that the occurrence of this scenario is due to the
existence of the stable and unstable manifolds, i.e., the dynamics
may follow the stable manifold (or the unstable manifold backward in time)
for almost infinite time.

In order to understand the
dispersive wave $\xi$ at the time $t_2$, we first notice that
the radiation  generated by the changes of
the masses of the bound states contribute to the dispersive wave.
We shall call it
the local part of the dispersive wave.
This local part, responsible for the relaxation of the excited states,
will always  be of the same order as the
main decay term and will not be small.
Our key observation is that the rest of
$\xi(t_2)$, call the global part,
is negligible when measured by an outgoing estimate.
To control the local part, we apply an initial layer
argument in the interval $[t_2, t_2+ \Delta t_2]$ until it becomes small
at the time $t_2+ \Delta t_2$.
Thus up to minor changes,
we can now apply the argument of \cite{TY2} from this time and
the dynamics will converge to some ground state.
The essence of this approach  is that  it  {\it extracts
the local relevant part of the dispersive wave  while  treating the global
part as an error term by measuring it with an outgoing estimate.}

The scheme we just described is for heuristic
explanation. Its precise form will be given  in section 3.
For nonlinear Schr\"odinger equations with general potentials,
the analysis will be more complicated.
In the two bound states case,
if the condition  $e_0<2 e_1$ fails,
the decay will be much slower than $1/\sqrt t$ and thus all errors have to be
controlled much more accurately. The picture is even more
complicated for multiple-bound states.
The resonance decay may be extremely slow (such as $t^{-\e}$);
the excited states may decay to other lower energy excited states
before finally decay to a ground state.
So far there has been no
rigorous work in this direction. However,
the notion of outgoing estimates and the initial layer argument seem to provide
the right general notion for estimating the dispersive wave.

\section{Preliminaries}

Denote $\bka{t}=1+|t|$ and $\bka{x}=1+|x|$. We define $L^2\loc$
and $L^1\loc$-norms by
\[
\norm{f}_{L^2\loc}=\norm{\bka{x}^{-r_1} f}_{L^2}, \qquad
\norm{f}_{L^1\loc}=\norm{\bka{x}^{-2r_1} f}_{L^1},
\]
where $r_1>3$ is a constant to be determined by \eqref{eq:22-1B}.

\subsection{Nonlinear bound states and linear decay estimates}

We recall some results for nonlinear bound states and linear
estimates from \cite{TY,TY2}.

\medskip

\begin{lemma} \label{th:2-1}
Suppose that $-\Delta+V$ satisfies the assumptions A0 and A2.
There is a small constant $ n_0>0$ such that the following hold.
For any $E$ between $e_0$ and $e_0+ \la n_0^2$ there is a
nonlinear ground state $Q_{E}$ solving \eqref{Q.eq}. The nonlinear
ground state $Q_{E}$ is real, local, smooth, $\la^{-1}(E-e_0)>0$,
and
\[
Q_{E} = n \,\phi_0 +h, \qquad h \perp \phi_0, \qquad
h=O(n ^{3 }),
\]
where $n = [(E-e_0)/(\la \int \phi_0^4 \, dx)]^{1/2}$.
Moreover, we have $R_E\equiv \pd_E Q_{E} = C n^{-2} \, Q_{E} + O(n )=
O( n^{-1} )$ and $\pd_E^2 Q_{E} = O( n^{-3} )$. If we define
$c_1\equiv (Q,R)^{-1}$, then $c_1=O(1)$ and $\la c_1
>0$.

There is also a family of nonlinear excited states
$\bket{Q_{E_1}}_{E_1}$ for $E_1$ between $e_1$ and $e_1+\la n_0^2$
satisfying similar properties: $Q_{E_1}=m \phi_1 + O(m^3)$ solves
\eqref{Q.eq} with $m\sim C [\la^{-1}(E_1-e_1)]^{1/2}$, etc.

\end{lemma}

\medskip

This lemma can be proven using standard perturbation argument, see
\cite{TY}. For the purpose of this paper, we prefer to use the
value $m=(\phi_1, \, Q_1)$ as the parameter and refer to the
family of excited states as $Q_1(m)$.

\medskip

\begin{lemma}[decay estimates for $e^{-itH_0}$] \label{th:2-2}
Suppose that $H_0=-\Delta+V$ satisfies the Assumptions A0--A2.
For $q \in [2, \infty]$ and
$q'=q/(q-1)$,
\begin{equation}  \label{eq:22-1A}
\norm{  e^{-itH_0} \, \Pc^{H_0} \phi }_{L^q} \le C \,|t|^ {-3
\bke{\frac 12 - \frac 1q}} \norm{\phi}_{L^{q'}}  .
\end{equation}
For sufficiently large $r_1$, we have
\begin{equation} \label{eq:22-1B}
\lim_{\sigma\to 0+} \norm{ \bka{x}^{-r_1} \, e^{-it H_0} \, \frac
1{H_0  + e_0 - 2 e_1- \sigma i} \Pc^{H_0} \bka{x}^{-r_1} \phi
}_{L^2} \le C \bka{t}^{-3/2} \norm{\phi}_{L^2}.
\end{equation}

\end{lemma}

\medskip

The decay estimate \eqref{eq:22-1A} is contained in  \cite{JSS}
and \cite{Y}; the estimate \eqref{eq:22-1B} is taken from
\cite{SW2} and  \cite{TY}. The estimate \eqref{eq:22-1B} holds
only if we take $\sigma\to 0+$, not $\sigma\to 0-$.

\subsection{Equations and decompositions}

For initial data near excited states, the decomposition
\eqref{psidec0} contains an error of order $y^3$ and it is
difficult to read from \eqref{psidec0} whether the wave function
is exactly an excited state.  Thus  we shall use the decomposition

\begin{equation}
\label{psidec1} \psi = {x}\phi_0 + Q_1(y) +{\xi}  ,
\end{equation}
where
\begin{equation}
y=\lbar{y}, \quad x =\lbar{x} -(\phi_0, Q_1(y)) ,\quad
\xi=\lbar{\xi}-\Pc Q_1(y)  .
\end{equation}
Here we have used the convention that
$$
Q_1(y) :=Q_1(m) e^{i \Theta}, \qquad m= |y|, \; \; m e^{i \Theta}
= y.
$$
For $\psi$ with sufficiently small $L^2$ norm, such a
decomposition exists and is unique \cite{TY2}. Thus we shall write
\begin{equation}\label{eq:2-4}
\psi(t) = x(t)\phi_0 + Q_1(m(t))e^{i \Theta(t)} + \xi(t)  ,\qquad
\xi(t) \in \Hc(H_0).
\end{equation}
%
If we write $\Theta(t)=\theta(t) -
\int_0^t E_1(m(s)) \, ds$, we can write $y(t)$ as
\begin{equation} \label{y.def}
y= m e^{i \Theta} = m \exp \bket{i\theta(t) - i\int_0^t E_1(m(s))
\, ds}  .
\end{equation}
Denote the part orthogonal to $\phi_1$ by $h ={x}\phi_0 +{\xi}$.
From the Schr\"odinger equation \eqref{Sch},  $h$ satisfies the equation
\begin{align}
i \pd_t h &= H_0 h + G + \Lambda, \nonumber
\\
G &= \la |\psi|^2 \psi - \la Q_1^3 e^{i \Theta} \nonumber \\
&=\la Q_1^2(e^{i2\Theta} \bar h +2 h ) + \la Q_1(e^{i\Theta}
2h\bar h +e^{-i\Theta} h^2 ) + \la |h|^2 h,
\\
\Lambda &=  \bke{\dot \theta Q_1 - i \dot m Q_1' } e^{i \Theta}
,\qquad \bke{Q_1'(m) := \tfrac d{dm} Q_1(m)}. \label{Lambda.def}
\end{align}
Since  $m(t)$ and $\theta(t)$ are chosen so that \eqref{eq:2-4}
holds, we have  $0 = (\phi_1, i\pd_t h(t)) = \bke{ \phi_1, \, G +
(\dot \theta Q_1 - i \dot m Q_1')e^{i \Theta} } $. Hence $m(t)$
and $\theta(t)$ satisfy
\begin{equation} \label{mt.eq}
\dot m = \bke{ \phi_1, \Im G e^{-i \Theta} } \, , \qquad \dot
\theta = -\frac 1m \, \bke{ \phi_1, \Re G e^{-i \Theta}}  .
\end{equation}

We also have the equation for $y$:
\[
i \dot y  = i \dot m e^{i \Theta} - ( \dot \theta- E_1(m) ) m e^{i
\Theta} = E_1(m)y + e^{i \Theta} (i \dot m - m \dot \theta ) =
E_1(m)y + (\phi_1 , G) .
\]
Here we have used \eqref{mt.eq}.  Denote $\Lambda_\pi= \pi
\Lambda$ where $\pi$ is the orthogonal projection $\pi \psi = \psi
- (\phi_1, \psi) \phi_1$. We can decompose the equation for $h$
into equations for $x$ and $\xi$. Thus  the original Schr\"odinger
equation is equivalent to
\begin{equation}
\label{xyxi.eq} \left \{
\begin{aligned}
i\dot x &= e_0 \, x + \bke{ \phi_0, \, G + \Lambda_\pi },
\\
i \dot y & = E_1(m)y + (\phi_1 , G),
\\
i \pd_t \xi &= H_0 \, \xi + \Pc \bke{ G + \Lambda_\pi }.
\end{aligned} \right .
\end{equation}

Clearly, $x$ has an oscillation factor $e^{-ie_0t}$, and $y$ has a
factor $e^{-ie_1t}$ since $E_1(m)\sim e_1$. Hence we define
\begin{equation} \label{eq:uv}
  x(t)= e^{-i e_0 t} u(t), \qquad y(t)=e^{-i e_1 t} v(t).
\end{equation}
Together with the integral form of the equation for $\xi$, we have
\begin{align}
\dot u &= -i e^{ie_0t}  \bke{ \phi_0, \, G + \Lambda_\pi }
,\label{u.eq}
\\
\dot v &= -i e^{ie_1 t} \bkt{(E_1(m)-e_1)y + (\phi_1 , G)},
\label{v.eq}
\\
\xi(t) &= e^{-i H_0  t}\xi_0  +\int_0^t e^{-i H_0 (t-s)} \, \PcH
G_\xi(s)\, d s  ,\quad G_\xi = i^{-1}(  G + \Lambda_\pi)
 .\label{xi.eq}
\end{align}
This is the system we shall study.

We denote by $G_3$ the leading terms of $G$, which consists of
cubic monomials in $x$ and $y$:
\begin{equation}\label{G3.def}
G_3 =  \la (y^2 \bar x + 2|y|^2 x) \phi_0 \phi_1^2 + \la (2|x|^2 y
+ x^2 \bar y) \phi_0^2 \phi_1 +\la |x|^2 x \phi_0^3 .
\end{equation}

We can expand  $E_1(m)$ in $m$ as
\begin{equation} \label{E.dec}
  E_1(m)= e_1 + E_{1,2} m^2 + E_{1,4}m^4 + E_1^{(6)}(m),
  \quad E_1^{(6)}(m)= O(m^6)  .
\end{equation}

We think of $x$ and $y$ as order $n$, and $\xi$ as order $n^3$.
%
Since, by \eqref{Lambda.def}--\eqref{mt.eq}, $\Lambda_\pi$ is
local and
\begin{equation}\label{Lambdapi-est}
\norm{\Lambda_\pi} \le |\dot \theta| \norm{\pi Q_1} + |\dot m|
\norm{\pi Q_1'} \le C |y|^2 \norm{G}\loc,
\end{equation}
the main terms in $G_\xi=i^{-1}(G+\Lambda_\pi)$ is $i^{-1}G_3$.
These terms are explicit and can be integrated. We integrate the
first term $\la y^2 \bar x \phi_0 \phi_1 ^2$ in $G_3$ as an
example:
\begin{align*}
& - i \la \int_0^t e^{-i H_0  (t-s)} \, \Pc y^2 \bar x \phi_0
\phi_1 ^2 \, d s
\\
&= - i \la e^{-i H_0  t} \int_0^t  e^{i (H_0 -0i) s} \, e^{i
(e_0-2e_1) s} \, v^2 \bar u \Pc \phi_0 \phi_1 ^2 \, d s
\\
&= y^2 \bar x \Phi_1 - e^{-i H_0  t} y^2 \bar x(0) \Phi_1 -
\int_0^t e^{-i H_0 (t-s) } \, e^{i (e_0-2e_1) s} \frac d{d s}
\bke{ v^2 \bar u} \Phi_1  \, d s  ,
\end{align*}
where
\begin{equation}\label{Phi1.def}
\Phi_1 = \frac {-\la} { H_0 -0i +e_0- 2e_1} \, \Pc \phi_0 \phi_1
^2  .
\end{equation}
This term,  with the phase factor  $e_0 - 2 e_1$,  is the only one
in $G_3$ having  a negative phase factor. 
Since $-(e_0 - 2 e_1)$ is in  the continuous spectrum of $H_0 $,
$H_0 +e_0- 2e_1$ is not invertible, and needs a regularization
$-0i$. We choose $-0i$, not $+0i$, so that the term $e^{-i H_0  t}
y^2 \bar x(0) \Phi_1$ decays as $t \to \infty$, see Lemma \ref{th:2-2}.

We can integrate all terms in $G_3$ and obtain  the main terms of
$\xi(t)$ as
\begin{equation}  \label{2-17}
\xi^{(2)}(t)= y^2 \bar x \Phi_1 + |y|^2 x \Phi_2 + |x|^2 y \Phi_3
+ x^2 \bar y \Phi_4  + |x|^2 x \Phi_5 ,
\end{equation}
where
\begin{alignat}{2} \label{Phij.def}
\Phi_2 &=  \frac {-2\la} { H_0  -e_0 } \, \Pc \phi_0 \phi_1 ^2  ,
\qquad &\Phi_3& =  \frac {-2\la} { H_0  -  e_1} \, \Pc \phi_0^2
\phi_1  ,
\\
\Phi_4 &=  \frac {-\la} { H_0   -2 e_0 + e_1} \, \Pc \phi_0 ^2
\phi_1  , \qquad &\Phi_5 &=  \frac {-\la} { H_0   - e_0} \, \Pc
\phi_0 ^3 . \nonumber
\end{alignat}
The rest of $\xi(t)$ is
\begin{align}
\xi^{(3)}(t)&= e^{-i H_0  t} \xi_0 -e^{-i H_0  t}\xi^{(2)}(0)
-\int_0^t e^{-i H_0 (t-s)} \, \Pc \, G_4 \, d s \nonumber
\\
&\quad + \int_0^t e^{-i H_0 (t-s)} \, \Pc
\bke{G_\xi-i^{-1}G_3-i^{-1} \la|\xi|^2\xi} \, d s \nonumber
\\
& \quad + \int_0^t e^{-i H_0 (t-s)} \, \Pc
\bke{i^{-1}\la|\xi|^2\xi}\, d s \nonumber
\\
&\equiv \xi^{(3)}_1(t)+ \xi^{(3)}_2(t) +\xi^{(3)}_3(t)
+\xi^{(3)}_4(t)+ \xi^{(3)}_5(t). \label{xi3.def}
\end{align}
The integrand $G_4$ in $\xi^{(3)}_3(t)$ consists of the remainders
from the integration by parts:
\begin{align}
G_4 &= e^{i(e_0 - 2 e_1)  s} \frac d{d s} \bke{ v^2 \bar u} \Phi_1
+ e^{i(-e_0 )  s} \frac d{d s} \bke{ |v|^2 u} \Phi_2
\label{G4.def}
\\
& \quad + e^{i ( -  e_1)  s} \frac d{d s} \bke{ |u|^2  v} \Phi_3 +
e^{i (-2 e_0 + e_1)  s} \frac d{d s} \bke{ u^2 \bar v} \Phi_4 +
e^{i (-e_0 )  s} \frac d{d s} \bke{ u^2 \bar u} \Phi_5 .\nonumber
\end{align}
The integrands of $\xi^{(3)}_4(t)$ and $\xi^{(3)}_5(t)$ are higher
order terms of $G_\xi$ which we did not integrate. We single out
$\xi^{(3)}_5(t)$ since $|\xi|^2\xi$ is a non-local term. Thus we
have the following decomposition for $\xi$:
\begin{equation}\label{xi.dec}
\xi(t)= \xi^{(2)}(t)+\xi^{(3)}(t) = \xi^{(2)} +\bke{\xi^{(3)}_1 +
\cdots + \xi^{(3)}_5}.
\end{equation}
Denote $\xi^{(3)}_{1-2} = \xi^{(3)}_1 + \xi^{(3)}_2$ and
$\xi^{(3)}_{3-5} = \xi^{(3)}_3 + \xi^{(3)}_4 + \xi^{(3)}_5$. We
have
\begin{equation}\label{xi3.dec}
\begin{split}
\xi^{(3)}_{1-2}(t) &= e^{-itH_0} [\xi_0 - \xi^{(2)}(0)], \\
\xi^{(3)}_{3-5}(t) &= \int_0^t e^{-i(t-s)H_0} \Pc (G_\xi -
i^{-1}G_3 - G_4)(s) ds.
\end{split}
\end{equation}
We now derive a bound for $\norm{\xi^{(3)}_{3-5}(t) }_{L^2\loc}$.
Using Lemma \ref{th:2-2} to estimate the integrand of $\xi^{(3)}_3$ and
bounding the $L^2\loc$-norm of the integrand of $\xi^{(3)}_4+\xi^{(3)}_5$
by either its $L^\infty$
or $L^4$-norm, we have, assuming \eqref{2-24} below,
\begin{equation}\label{xi335.est}
\norm{\xi^{(3)}_{3-5}(t) }_{L^2\loc} \le \int_0^t \min
\bket{|t-s|^{-3/2},|t-s|^{-3/4}} g_{\xi,3-5}(s) ds,
\end{equation}
where
\begin{equation}\label{gxi35.def}
g_{\xi,3-5}(t) \equiv C \norm{ G_\xi - i^{-1}G_3 }_{L^1 \cap
L^{4/3}} + C n^2 |\dot u|+ C n |u \dot v| .
\end{equation}

\bigskip

\begin{lemma} \label{th:Gxi}
Suppose
\begin{equation}\label{2-24}
|x|,|y|\le n \le \nxi \ll 1, \quad \norm{\xi}_{L^2\cap L^4} \le
\nxi.
\end{equation}
Denote $X = n \nxi \norm{\xi}_{L^2\loc} + \nxi \norm{\xi}_{L^4}^2$.
We have
\begin{align}
\norm{G}_{L^1 \loc}+\norm{G_\xi(t)}_{L^1 \cap L^{4/3}} &\lesssim
n^2 x + X, \label{Gxi-est1}
\\
\norm{G-G_3}_{L^1 \loc} + g_{\xi,3-5}(t) &\lesssim n^4 x + X.\label{Gxi-est2}
\end{align}
\end{lemma}

\bigskip

\myproof From the definitions of $G,G_3$ and by H\"older
inequality, we have
\begin{equation*}
 \norm{G- G_3}_{L^1 \cap L^{4/3}} \lesssim (*),
\end{equation*}
where
\[
(*) = n^4 x + n^2 \norm{\xi}_{L^2\loc} + n \norm{\xi}_{L^2 \cap
L^4} \norm{\xi}_{L^2\loc} + \norm{\xi}_{L^2 \cap L^4}
\norm{\xi}_{L^4}^2.
\]
We have $\norm{G-G_3}_{L^1 \loc}\les \norm{G- G_3}_{L^1 \cap L^{4/3}} \les (*)$
and $\norm{G}_{L^1 \loc}\les \norm{G_3}_{L^1 \loc} + \norm{G-G_3}_{L^1 \loc} \les
n^2 x+ (*)$. Since $G_\xi = i^{-1}(G+\Lambda_\pi)$ with
$\norm{\Lambda_\pi} \le n^2 \norm{G}_{L^1 \loc}$ by
\eqref{Lambdapi-est}, we have
\[
\norm{ G_\xi - i^{-1}G_3 }_{L^1 \cap L^{4/3}} \lesssim (*), \qquad
\norm{ G_\xi}_{L^1 \cap L^{4/3}} \lesssim n^2x + (*).
\]
By \eqref{u.eq}--\eqref{v.eq}, also using \eqref{Lambdapi-est},
\begin{equation}\label{dotuv-est}
|\dot u| \les \norm{G}_{L^1 \loc} + \norm{\Lambda_\pi}_{L^1 \loc}
\les  \norm{G}_{L^1 \loc} ,\qquad |\dot v| \les \norm{G}_{L^1
\loc} + n^3.
\end{equation}
From the definition \eqref{gxi35.def} of $g_{\xi,3-5}(t)$,
\eqref{dotuv-est} and
$\norm{G}_{L^1 \loc}\les n^2 x+ (*)$, we have
\[
g_{\xi,3-5}(t) \lesssim (*) +  n^2 \norm{G}_{L^1 \loc} + n |u| n^3
\les (*).
\]
From the assumption \eqref{2-24}, $(*) \lesssim n^4 x + n \nxi
\norm{\xi}_{L^2\loc} + \nxi \norm{\xi}_{L^4}^2$. Thus we have
proved the Lemma. \myendproof

\newpage 

\subsection{Normal form for equations of bound states}

Recall that we write $x(t)= e^{-ie_0t} u(t)$ and $y(t)= e^{-ie_1t}
v(t)$. We have the following normal form for the equations of
$\dot u$ and $\dot v$.

\bigskip

\begin{lemma}[Normal form] \label{th:NF}
Suppose
\begin{equation} \label{2-21}
|x(t)|,|y(t)|\le n \ll 1, \qquad \norm{\xi(t)}_{L^2 \cap L^4} \ll
1.
\end{equation}
There are perturbations $\mu$ of $u$ and $\nu$ of $v$ satisfying
\begin{equation}
|u(t)-\mu(t)|+|v(t)-\nu(t)| \le C_1\, n^2|x(t)|, \label{eq:33-1}
\end{equation}
such that
\begin{equation} \label{eq:NF}\begin{split}
\dot \mu  &= (c_1 |\mu|^2 + c_2|\nu |^2) \mu  + (c_3 |\mu|^4 + c_4
|\mu |^2|\nu|^2 + c_{5} |\nu |^4)\mu  + g_u ,
\\
\dot \nu &= (c_6 |\mu|^2 + c_7 |\nu |^2) \nu + (c_8 |\mu|^4 + c_9
|\mu |^2|\nu|^2 + c_{10}|\nu |^4)\nu + g_v .
\end{split}\end{equation}
Here $g_u$ and $g_v$ are error terms. All coefficients $c_1,
\cdots, c_{10}$ are of order one and, except $c_5$ and $c_9$,
purely imaginary. We have
\begin{equation}
\Re c_5 = \gamma_0  , \qquad \Re c_9 =  -2\gamma_0 ,
\end{equation}
where $\gamma_0>0$ is defined in \eqref{gamma0.def}.
Moreover, we can write $g_v$ as
\begin{equation} \label{gv.dec}
g_v = - i E^{(6)}(|y|)\nu + \wt g_v ,
\end{equation}
where $E^{(6)}(|y|) = O(|y|^6)$ is defined in \eqref{E.dec},
and 
\begin{equation} \label{guv.est}
\begin{split}
&|g_u(t)| + |\wt g_v(t)|
 \\
& \le C_1 \bket{ \nxi n^5 |x| + n^2 \norm{\xi^{(3)}}_{L^2\loc} + n
\norm{\xi}_{L^2\loc}^2 + \bke{\norm{\xi}_{L^2\loc} +\nxi n^2}
\norm{\xi}_{L^4}^2  },
\end{split}
\end{equation}
for some explicit constant $C_1$.
\end{lemma}

\bigskip

\myproof This is Lemma 3.4 of \cite{TY2}. The definitions of
$\mu,\nu,g_u,g_v$ are exactly the same. The only difference is the
error estimates \eqref{guv.est} since our assumption \eqref{2-21}
is different from that in \cite{TY2}. Since $\mu$ is of the form
$u + n^2 u + n^4 u$ and $\nu$ of the form $v + n^2 u + n^4 u$,
their estimates remain the same. We only need to prove
\eqref{guv.est}.

In \cite{TY2} $g_u$ and $g_v$ are defined as
\begin{align*}
g_u &= g_{u,4} + g_{u,5}  + g_{u,3}     + R_{u,7} -i  e^{i e_0 t}
(\phi_0,\,  G_{5,3}) ,
\\
g_v&= g_{v,4} +g_{v,5} + g_{v,3} + R_{v,7} -i  e^{i e_1 t}
(\phi_1,\,  G_{5,3}). 
\end{align*}
See \cite{TY2} for their exact definitions. Recall from
\cite{TY2} that $g_{u,3}$ consists of
higher order terms of $g_{u,1}$ and $g_{u,2}$. Note $g_{u,1}$
consists of terms of the form $n^2 \dot u + n u \dot v$, and
$g_{u,2}$ consists of terms of the form $n^2 (u-\mu) + n u (v -
\nu)$. Together with \eqref{u.eq}, \eqref{v.eq} and
\eqref{Lambdapi-est},  we can bound
$g_{u,3}$ by
\[
|g_{u,3}|\les n^2 \norm{G-G_3}_{L^1\loc} +n^4 \norm{G}_{L^1\loc} +
n^6 |x|.
\]
The other terms in $g_u$ are of the form:
\begin{align*}
g_{u,4}& = n^4 \dot u + n^3 u \dot v,\\
g_{u,5} &= n^4 (u-\mu) + n^3 u (v - \nu),\\
R_{u,7} &=(\phi_0,\, n^6x+n^4 \xi + n\xi^2 + \xi^3
+ n^2 \norm{G-G_3}_{L^1\loc} + n^4 \norm{G}_{L^1\loc}),\\
G_{5,3}&= n^2 \xi^{(3)}.
\end{align*}
We can bound $\dot u$, $\dot v$, $\norm{G}_{L^1\loc}$ and
$\norm{G-G_3}_{L^1\loc}$ by \eqref{dotuv-est}, \eqref{Gxi-est1} and
\eqref{Gxi-est2}.
Summing the estimates, we have
\begin{align*}
|g_u|&\les n^2 \norm{G-G_3}_{L^1\loc}  + n^4 \norm{G}_{L^1\loc}
+n^6 x + n^4 \xi + n\xi^2 + \xi^3 + n^2 \xi^{(3)}
\\
& \les n^6 x + n^2 \xi^{(3)} +n^4 \xi + n\xi^2 + \xi^3 +  n^2 X,
\end{align*}
where  $X = n \nxi \norm{\xi}_{L^2\loc} + \nxi \norm{\xi}_{L^4}^2$
and all terms with $\xi$ are measured in $L^1\loc$. Therefore
\[
|g_u(t)| \les  n^6 |x| + n^2 \norm{\xi^{(3)}}_{L^2\loc} + n^3 \nxi
\norm{\xi}_{L^2\loc} + n \norm{\xi}_{L^2\loc}^2 +
\bke{\norm{\xi}_{L^2\loc}+ n^2\nxi}\norm{\xi}_{L^4}^2 .
\]
Note $\norm{\xi}_{L^2\loc} \les n^2 x+\norm{\xi^{(3)}}_{L^2\loc}$.
Hence we obtain the estimate of $g_u$ in \eqref{guv.est}. The
estimate of $\wt g_v$ is proved in the same way. \myendproof

As a result of the lemma, we have
\begin{equation} \label{|mu|}
\frac d{ dt} |\mu| = \frac 1{2|\mu|}  \frac d{ dt} |\mu|^2 =
|\mu|^{-1} \Re \bar \mu \dot \mu = \gamma_0 |\nu|^4 |\mu|+ \Re
g_\mu \bar \mu /|\mu|.
\end{equation}
Similarly, using $\Re g_v \bar \nu = \Re \wt g_v \bar \nu$,
\begin{equation} \label{|nu|}
\frac d{ dt} |\nu| = -2 \gamma_0 |\mu|^2 |\nu|^3  + \Re \wt g_\nu
\bar \nu /|\nu|.
\end{equation}

\subsection{Relaxation to Ground States}

We shall need Theorem 4.3 of \cite{TY2} which provides a
relaxation estimates to ground states from initial data near some
ground state. It is a strengthened form of Theorem~1.3 in
\cite{TY}. For the purpose of the application in this paper, we
start the dynamics at $t=t_4$.

\begin{theorem} [\cite{TY2}] \label{th:1-1}
There are small constants $n_0,\e_0>0$ such that the following
hold. Suppose $\psi(t_4) = x(t_4) \phi_0 + Q_1(y(t_4)) +
\xi(t_4)$ with
\[
|x(t_4)| = n \ll n_0, \quad |y(t_4)| \le \e_0 n,
\]
and that $\xi(t_4)$ satisfies, for all $s\ge 0$,
\begin{equation}\label{outgoingest}
\begin{split}
\norm{\xi(t_4)}_{H^1} &\ll 1 ,
\\
\norm{ e^{-isH_0} \xi(t_4)}_{L^4} & \le C n^3 \dt (\dt + s)^{-3/4}
,
\\
\norm{ e^{-isH_0} \xi(t_4)}_{L^2 \loc} & \le C n^3 \frac
{\dt}{\dt+s}\, (1+s)^{-1/2},
\end{split}
\end{equation}
for some $\dt \in [1, n^{-4-1/4}]$. Then there is a frequency
$E_\infty$ and a function $\Theta(t)$ such that
$\norm{Q_{E_\infty}}_Y \sim n$, $\Theta(t) = -E_\infty t+
O(\log t)$ and, for some constant $C_2$,
\[
 \norm{\psi(t) - Q_{E_\infty}e^{i \Theta(t)} }_{L^2 \loc}
\le C_2 \bke{(\e n)^{-2}+ \gamma_0 n^2 (t-t_4)} ^{-1/2}.
\]
\end{theorem}

\subsection{Inequalities}

For convenience of reference, we collect some integral
inequalities here.

\bigskip

\begin{lemma} \label{th:cal}
(1) Suppose $t \ge T$, $\dt \ge 1$.
\begin{equation}\label{cal-1}
\int _{T-\dt}^{T} |t-s|^{-3/4} \, d s \le C \dt (\dt +
t-T)^{-3/4}.
\end{equation}
\begin{equation}  \label{cal-2} \int_{T-\dt}^{T}
\min \bket{(t-s)^{-3/2}, \ (t-s)^{-3/4} }  \, d s \le C \frac
{\dt}{\dt+t-T} \bka{t-T}^{-1/2}.
\end{equation}

(2) For $t\ge T \ge 1$,
\begin{equation}  \label{cal-3} \int_{T}^{t} (t-s)^{-3/4} \,
s^{-3/2} \, d s \le  C T^{-1/2} t^{-3/4}.
\end{equation}
\begin{equation}  \label{cal-4} \int_{T}^{t} \min \bket{(t-s)^{-3/2}, \
(t-s)^{-3/4} } \, s^{-3/2} \, d s \le C t^{-3/2}.
\end{equation}

\end{lemma}
\bigskip

\myproof (1) Let $t=T+\tau$. If $\tau > \dt$, $ (\dt + \tau) \sim
\tau$ and the integral in \eqref{cal-1} is bounded by $\int
_{T-\dt}^{T} \tau^{-3/4} \, d s = C \tau^{-3/4}\dt \sim C \dt (\dt
+ \tau)^{-3/4} $. If $\tau<\dt$, $ (\dt + \tau) \sim \dt$ and the
integral is bounded by $\int _{T-\dt}^{T} |T -s |^{-3/4} \, d s
= C (\dt)^{1/4} \sim C \dt (\dt + \tau)^{-3/4}$.

For \eqref{cal-2}, if $t \le T+1$, we have LHS $\le$ constant
$\le$ RHS. Hence we assume $t\ge T+1$. By a translation, \eqref{cal-2}
is equivalent to $\int_0 ^{\dt} (t-s)^{-3/2} ds
\le C (\dt) t^{-1}\bka{t-\dt}^{-1/2}$. The integral is bounded by
\begin{align*}
\int_0 ^{\dt} (t-s)^{-3/2} ds &= 2(t-\dt)^{-1/2} - 2t^{-1/2}
\\
&= 2 [(t-\dt)^{-1/2} + t^{-1/2}]^{-1}\, [(t-\dt)^{-1} - t^{-1}]
\\
&\le 2 (t-\dt)^{1/2} \bkt{(t-\dt)^{-1} t^{-1} \dt}.
\end{align*}


(2) Note, by rescaling,
\[
\int _{t/2}^{t} (t-s)^{-3/4} \, s^{-3/2} \, d s = C t ^{-5/4} \le
C T^{-1/2} t^{-3/4}.
\]
If $t< 2T$, the integral on the left of \eqref{cal-3} is bounded
by the above integral. If $t \ge 2T$, it is bounded by the sum of
the above integral and
\[
\int _{T}^{t/2} (t-s)^{-3/4} \, s^{-3/2} \, d s \le C t^{-3/4}
\int _{T}^{t/2} \, s^{-3/2} \, d s \le C t^{-3/4} T^{-1/2}.
\]
Hence \eqref{cal-3} is proven. For \eqref{cal-4}, note the left
side is bounded by
\[
\int _0^{t} \bka{t-s}^{-3/2} \bka{s}^{-3/2} \, d s +
\int_{t-1/2}^t (t-s)^{-3/4} t^{-3/2} d s,
\]
and both integrals are bounded by $C t^{-3/2}$. \myendproof

\newpage

\section{Estimates}

We have assumed that the initial data is small in $\| \cdot \|_Y$
in Theorem \ref{th:1-2}. We shall however use only the following
properties: Let $\psi_0 = x_0 \phi_0 + Q_1(y_0) + \xi_0$ with
$\xi_0 \in \Hc$. 
Then we have for all $t \ge 0$,
\begin{align}\label{1-11}
&|x_0| + |y_0| + \norm{\xi_0}_{L^2} \le \nxi , \nonumber
\\
&\norm{e^{-itH_0} \xi_0}_{L^4} \le \nxi \bka{t}^{-3/4},
\\
&\norm{e^{-itH_0} \xi_0}_{L^2 \loc }
\le \nxi \bka{t}^{-3/2}. \nonumber
\end{align}
From now on, we shall use these three conditions as our assumption
for Theorem \ref{th:1-2}.

Recall the orthogonal decomposition \eqref{psidec0} that
$\psi(t)=\lbar{x}\phi_0 + \lbar{y}\phi_1 + \lbar{\xi}$. We have
$|\lbar{x}(t)|^2 + |\lbar{y}(t)|^2 + \norm{\lbar{\xi}(t)}_{L^2}^2
=\norm{\psi(t)}_{L^2}^2 \le \nxi^2$. If we decompose $\psi(t)$ via
\eqref{psidec1}, i.e.,
\begin{equation}
\psi(t)=x\phi_0 + Q_1(y) + \xi  ,
\end{equation}
we have $y=\lbar{y}$, $x=\lbar{x}+O(y^3)$ and
$\xi=\lbar{\xi}+O(y^3)$. Thus
\begin{equation}  \label{eq:3-2}
|x(t)|, \; |y(t)|, \; \norm{\xi(t)}_{L^2} \le \tfrac 54 \nxi .
\end{equation}


Choose $\iota$ and $\delta$ so that
\begin{equation} \label{delta-sigma}
0 < \iota<0.2,\quad  0.6<\delta < 1, \quad \delta+\iota<1.
\end{equation}
(We set $\delta=3/4$ in the statement of case I in Theorem \ref{th:1-2}.)
%

Let
\begin{equation}
t_1\equiv \sup \bket{t\ge 0:\ \bke{ \max \bket{|x(s)|,|y(s)|}
}^{2+\delta} \le \nxi \bka{s}^{-3/2}, \ \forall s \in [0, t]}.
\end{equation}
$t_1$ may be $\infty$;
%
we may assume $t_1 \ge 1$ by enlarging $\nxi$.
Our guiding principle is the following chart. The time $t_2$ is defined in
Proposition~\ref{th:t2}.

\bigskip

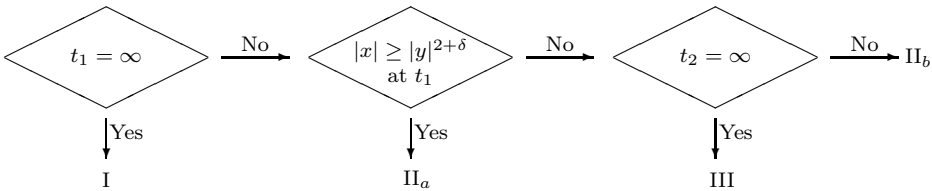
\setlength{\unitlength}{0.9mm}\noindent
\begin{picture}(105,30)

\put (0,20){\line(2,1){15}} \put (0,20){\line(2,-1){15}} \put
(30,20){\line(-2,1){15}} \put (30,20){\line(-2,-1){15}}

\put (32,20){\vector(1,0){10}} \put (15,11){\vector(0,-1){6}}
\put(15,2){\makebox(0,0)[c]{\scriptsize I}}
\put(15,20){\makebox(0,0)[c]{\scriptsize $t_1=\infty$}}
\put(18,9){\makebox(0,0)[c]{\scriptsize Yes} }
\put(37,22){\makebox(0,0)[c]{\scriptsize No} }

\put (45,20){\line(2,1){15}} \put (45,20){\line(2,-1){15}} \put
(75,20){\line(-2,1){15}} \put (75,20){\line(-2,-1){15}}

\put (77,20){\vector(1,0){10}} \put (60,11){\vector(0,-1){6}}
\put(61,2){\makebox(0,0)[c]{\scriptsize II$_a$}}
\put(60,19){\makebox(0,0)[c]{\scriptsize
$\begin{array}{c} |x|\ge |y|^{2+\delta}\\  \text{at }t_1\end{array}$}}
\put(63,9){\makebox(0,0)[c]{\scriptsize Yes} }
\put(82,22){\makebox(0,0)[c]{\scriptsize No} }

\put (90,20){\line(2,1){15}} \put (90,20){\line(2,-1){15}} \put
(120,20){\line(-2,1){15}} \put (120,20){\line(-2,-1){15}}

\put (122,20){\vector(1,0){10}} \put (105,11){\vector(0,-1){6}}
\put(106,2){\makebox(0,0)[c]{\scriptsize III}}
\put(105,20){\makebox(0,0)[c]{\scriptsize $t_2 =\infty$}}
\put(108,9){\makebox(0,0)[c]{\scriptsize Yes} }
\put(127,22){\makebox(0,0)[c]{\scriptsize No} }

\put(135,20){\makebox(0,0)[c]{\scriptsize II$_b$}}

\end{picture}

{\scriptsize
\begin{enumerate}
\item[I.] $\psi(t)$ vanishes locally.

\item[II$_a$.]  $\psi(t)$ relaxes to a ground state and stays away from nonlinear excited states
for all time.

\item[II$_b$.]  $\psi(t)$ approaches  some nonlinear excited state but
then relaxes to a ground state.

\item[III.]  $\psi(t)$ converges to a nonlinear excited state.
\end{enumerate}
}

The analysis of case II$_b$ is very subtle since the time scale
that $\psi(t)$ stays near an excited state may be infinite
compared to its local size. We have the following time line
picture for this case:

%
%
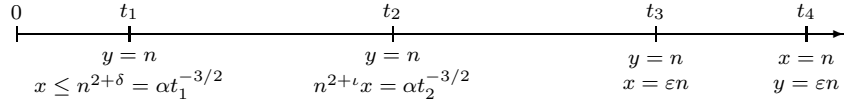
\setlength{\unitlength}{1mm}
\begin{picture}(105,15)
\put(0,10){\vector(1,0){110}}

\put (0,9){\line(0,1){2}} \put (0,13){\makebox(0,0)[c]{\scriptsize
$0$}}

\put (15,9){\line(0,1){2}} \put
(15,13){\makebox(0,0)[c]{\scriptsize $t_1$}} \put
(10,0){\makebox(10,10)[c]{\scriptsize $\begin{array}{c} y=n\\
 x \le n^{2+\delta} = \nxi t_1^{-3/2} \end{array}$}}

\put (50,9){\line(0,1){2}} \put
(50,13){\makebox(0,0)[c]{\scriptsize $t_2$}} \put
(45,0){\makebox(10,10)[c]{\scriptsize $\begin{array}{c} y=n\\ n
^{2+\iota} x = \nxi t_2^{-3/2} \end{array}$}}


\put (85,9){\line(0,1){2}} \put
(85,13){\makebox(0,0)[c]{\scriptsize $t_3$}} \put
(80,0){\makebox(10,10)[c]{\scriptsize $\begin{array}{c} y=n \\ x =
\e n \end{array}$}}

\put (105,9){\line(0,1){2}} \put
(105,13){\makebox(0,0)[c]{\scriptsize $t_4$}} \put
(100,0){\makebox(10,10)[c]{\scriptsize $\begin{array}{c} x=n \\ y
= \e n \end{array}$}}

\end{picture}

We first establish an estimate in the interval $[0, t_1)$.

\begin{proposition} \label{th:t1}
For $t\in [0,t_1)$, we have
\begin{equation} \label{t1-est}
\begin{split}
|x(t)|,|y(t)| &\le [\nxi \bka{t}^{-3/2}]^{1/(2+\delta)},
\\
\norm{\xi(t)}_{L^4} &\le (1+\iota) \ \nxi \bka{t}^{-3/4},
\\
\norm{\xi(t)}_{L^2\loc} &\le (1+\iota) \ \nxi \bka{t}^{-3/2},
\\
\norm{G_\xi(t)}_{L^1 \cap L^{4/3}} &\le  C \nxi^{3/(2+\delta)}
\bka{t}^{-3/2}.
\end{split}
\end{equation}

Suppose $ t_1= \infty$. Then
\begin{equation} \label{3-8}
|x(t)| + |y(t)| + \norm{\xi(t)}_{L^2\loc} \le C
t^{-3/(4+2\delta)}, \qquad \text{as } t \to \infty.
\end{equation}

Suppose $ t_1< \infty$. Let
\begin{equation}
n\equiv \max \bket{|x(t_1)|,\, |y(t_1)|}.
\end{equation}
We have $0< n < 2 \nxi$ and
\begin{equation} \label{3-5}
n^{2+\delta}= \nxi \bka{t_1}^{-3/2}.
\end{equation}
Moreover, for all $t \ge t_1$, we have the following outgoing estimates
on the dispersive wave $\xi$:
\begin{equation}\label{out:t1}
\begin{split}
\norm{e^{-i(t-t_1) H_0}\xi(t_1)}_{L^4} &\le (1+\iota) \  \nxi
t^{-3/4},
\\
\norm{e^{-i(t-t_1) H_0}\xi(t_1)}_{L^2\loc} &\le  (1+\iota) \  \nxi
t^{-3/2}.
\end{split}
\end{equation}

\end{proposition}

%

\bigskip

\myproof The estimate for $|x(t)|,|y(t)|$ in \eqref{t1-est}$_1$ is
by the definition of $t_1$. We will prove the rest of
\eqref{t1-est} by a continuity argument and assume that
\begin{equation} \label{3-1-3}
\norm{\xi(t)}_{L^4} \le 2 \nxi \bka{t}^{-3/4}, \qquad
\norm{\xi(t)}_{L^2\loc} \le 2 \nxi \bka{t}^{-3/2}.
\end{equation}
We explain the idea of continuity argument: Suppose the estimates
in \eqref{t1-est} is true only up to $t\le T$ with $T<t_1$. Since
the estimates \eqref{3-1-3} are weaker than those in
\eqref{t1-est}, they remain true for $t \in [0,T+\tau]$ for some
$\tau>0$, $T+\tau \le t_1$, by continuity. Our proof then implies
\eqref{t1-est} for  $t \in [0,T+\tau]$. This is a contradiction to
the choice of $T$. Hence \eqref{t1-est} holds for all $t \in
[0,t_1]$. We will use similar continuity arguments to prove
Propositions \ref{th:t2}--\ref{th:t4}.

Recall
\begin{equation}\label{3-1-xi}
\xi(t) = e^{-it H_0} \xi_0 + \int_0^t e^{-i(t-s) H_0} \Pc G_\xi(s)
\, d s ,
\end{equation}
and $G_\xi = i^{-1}(G+\Lambda_\pi)$. Since
\begin{equation} 
\norm{\xi^2 \bar \xi}_{L^{4/3}}\le \norm{\xi}_{L^{4}}^3,\qquad
\norm{\xi^2 \bar \xi}_{L^{1}}\le \norm{\xi}_{L^{2}}
\norm{\xi}_{L^{4}}^2,
\end{equation}
and $\norm{\xi}_{L^{4}\cap L^2}\le 2\nxi$, assuming \eqref{3-1-3}
we have
\begin{align*}
\norm{G_\xi(s)}_{L^1 \cap L^{4/3}} &\le C
(|x(s)|+|y(s)|)^{3/(2+\delta)} +  \norm{\xi(s)}_{L^4\cap L^2}
\norm{\xi(s)}_{L^4}^2
\\
&\le C (\nxi (1+s)^{-3/2})^{3/(2+\delta)} + C \nxi^3 (1+s)^{-3/2}
\\
&\le C \nxi^{3/(2+\delta)} (1+s)^{-3/2} = o(1) \ \nxi (1+s)^{-3/2}
.
\end{align*}
Here we have used $\delta < 1$, see \eqref{delta-sigma}. Using
\eqref{1-11} and \eqref{cal-3}, $\xi(t)$ is bounded in $L^4$ by
\[
\norm{\xi(t)}_{L^4} \le \nxi \bka{t}^{-3/4}
+ \int_0^{t}
C(t-s)^{-3/4} \norm{G_\xi(s)}_{ L^{4/3}} \, d s \le \frac
{(1+\iota) \nxi}{(1+t)^{3/4}}.
\]

To bound $ \norm{\xi(t)}_{L^2\loc}$, we bound the integrand in
\eqref{3-1-xi} in ${L^{\infty}}$ for $s$ small and in ${L^{4}}$
for $s$ large. Hence, using \eqref{1-11} and \eqref{cal-4},
\begin{align*}
\norm{\xi(t)}_{L^2\loc} &\le \nxi \bka{t}^{-3/2}
+ \int_0^{t} C\min \bket{ (t-s)^{-3/2} , (t-s)^{-3/4}}
\norm{G_\xi(s)}_{ L^{1}\cap L^{4/3}} \, d s \\
& \le (1+\iota) \nxi \bka{t}^{-3/2}.
\end{align*}
Hence we have shown all estimates in \eqref{t1-est}, by a
continuity argument.

Suppose $t_1 = \infty$. It follows from \eqref{t1-est} that
everything vanishes and we have \eqref{3-8}.
Suppose $t_1<\infty$. That $n\le 2 \nxi$ is by \eqref{eq:3-2}.
Eq.~\eqref{3-5} is by the definition of $t_1$. We want to show
\eqref{out:t1}. 
For $t \ge t_1$ we have
\begin{equation*}
e^{-i(t-t_1) H_0}\xi(t_1) = e^{-i t H_0} \xi_0 + \int_0^{t_1}
e^{-i(t -\tau) H_0} \Pc G_\xi(\tau) \, d \tau  .
\end{equation*}
Hence, using $\norm{G_\xi(\tau)}_{ L^{1}\cap L^{4/3} } \le o(1)
\nxi (1+s)^{-3/2}$ and \eqref{cal-3}--\eqref{cal-4}, we have
\begin{align*}
\norm{e^{-i (t-t_1) H_0}\xi(t_1)}_{L^4} &\le {\nxi}t^{-3/4} +
\int_0^{t_1} (t -\tau)^{-3/4}  \norm{G_\xi(\tau)}_{ L^{4/3}} \,
d \tau 
\\
& \le \nxi t^{-3/4} + o(1) \nxi t^{-3/4}
\le (1+\iota) \nxi t^{-3/4}, 
\end{align*}
\begin{align*}
& \norm{e^{-i (t-t_1) H_0}\xi(t_1)}_{L^2\loc} \\
& \quad\le {\nxi} t^{-3/2}
 + \int_0^{t_1} \min \bket {(t-\tau)^{-3/2},
(t-\tau)^{-3/4}} \norm{G_\xi(\tau)}_{ L^{1}\cap L^{4/3} } \, d
\tau 
\\
&\quad \le (1+\iota) \nxi t ^{-3/2} .
\end{align*}
This proves \eqref{out:t1} and we conclude the proof of Proposition \ref{th:t1}.
\myendproof

The significance of $t_1$ is that it is a time when the dispersion
loses its dominance over the bound states. If $t_1=\infty$, the
dispersion dominates for all the time and everything vanishes
locally by \eqref{3-8}.  This gives us case I of Theorem
\ref{th:1-2}. Suppose now $t_1<\infty$. There are two
possibilities:
\begin{align*}
1. & \quad |x(t_1)| \ge |y(t_1)|^{2+\delta} ,
\\
2. & \quad |x(t_1)| <|y(t_1)|^{2+\delta} . 
\end{align*}
%
We will focus on the second case since it is more subtle. We will
come back to the first case, which corresponds to case II$_a$, at
the end.

\begin{proposition} \label{th:t2}
Suppose $t_1<\infty$ and
\begin{equation} \label{3-15}
|y(t_1)|=n, \quad |x(t_1)|\le n^{2+\delta}, \quad \nxi t_1^{-3/2}
=  n^{2+\delta}.
\end{equation}
Define
\begin{equation}
t_2\equiv \sup \bket{t\ge t_1:\ n^{2+\iota} |x(s)| \le \nxi
\bka{s}^{-3/2} , \ \forall s \in [t_1,t]}.
\end{equation}
For $t\in [t_1,t_2)$, we have
\begin{equation} \label{t2-est}
\begin{split}
|y(t)/y(t_1)| &\in [\tfrac 78,\ \tfrac 98],
\\
|x(t)| &\le  \min \bket{ 2 n^{2+\delta}, n^{-2-\iota} \nxi
\bka{t}^{-3/2}},
\\
\norm{\xi(t)}_{L^4} &\le (1+2\iota)  \nxi \bka{t}^{-3/4},
\\
\norm{\xi(t)}_{L^2\loc} &\le C n^{-\iota} \nxi \bka{t}^{-3/2}.
\end{split}
\end{equation}

Suppose $t_2=\infty$. Then there is a $ y_\infty \sim n$ such that
\begin{equation} \label{3-18}
\big| |y(t)| - y_\infty \big|+ |x(t)| + \norm{\xi(t)}_{L^2\loc}
\le C t^{-1/2} ,\qquad \text{as } t \to \infty.
\end{equation}
Moreover, $\Theta(t) = - E_1(y_\infty) t + O(t^{1/2})$ as $t \to
\infty$, where $\Theta(t)$ is the phase of $y(t)$, defined in
\eqref{y.def}.

Suppose $t_2 < \infty$. We have $n^{2+\iota} |x(t_2)| = \nxi
\bka{t_2}^{-3/2} $ and, for all $t \ge t_2$,
the following outgoing estimates
on the dispersive wave $\xi$:
\begin{equation}\label{out:t2}
\begin{split}
\norm{e^{-i(t-t_2) H_0}\xi(t_2)}_{L^4} &\le  (1+2\iota) \nxi
t^{-3/4},
\\
\norm{e^{-i(t-t_2) H_0}\xi(t_2)}_{L^2\loc} &\le   (1+2\iota) \nxi
t^{-3/2} +  \frac {C n^{2}|x(t_2)| t_1}{t_1+ t-t_2}
\bka{t-t_2}^{-1/2}.
\end{split}
\end{equation}

\end{proposition}

\myproof We first consider $t \in [t_1, t_2)$. By definition of
$t_2$, we have
\begin{equation} \label{3-22}
|x(t)| \le n^{-2-\iota} \nxi t^{-3/2}, \qquad (t_1 \le t < t_2).
\end{equation}
Using a continuity argument we may assume
\begin{equation} \label{3-14}
\begin{split}
&|y(t)/y(t_1)| \in [\tfrac 12,\ \tfrac 32], \qquad |x(t)| \le 3
n^{2+\delta},
\\
&\norm{\xi(t)}_{L^4} \le  2 \nxi \bka{t}^{-3/4} , \qquad
\norm{\xi(t)}_{L^2\loc} \le 2C n^{-\iota} \nxi \bka{t}^{-3/2}.
\end{split}
\end{equation}

We first estimate $\xi(t)$. By Lemma \ref{th:Gxi},
$\norm{G_\xi(s)}_{L^1\cap L^{4/3}} \lesssim n^2 x + X$ with
\[
X(s) = n \nxi \norm{\xi}_{L^2\loc} + \nxi \norm{\xi}_{L^4}^2 \le
C(n^{1-\iota}\nxi^2 + \nxi^3) s^{-3/2},
\]
where we have used \eqref{3-14} in the last inequality.  Using \eqref{3-22},
we thus have $\norm{G_\xi(s)}_{L^1\cap L^{4/3}} \le Cn^{-\iota}\nxi s^{-3/2}$.
For $\xi(t)$ with $t \in [t_1,t_2)$ we have
\[
\xi(t) = e^{-i (t-t_1) H_0} \xi(t_1) + J(t), \qquad J(t)\equiv
\int_{t_1}^t e^{-i (t-s) H_0} \PcH G_\xi(s) \, d s.
\]
The estimate for $e^{-i (t-t_1) H_0} \xi(t_1)$ is provided by
\eqref{out:t1} of Proposition \ref{th:t1}. Hence it suffices to
estimate the integral $J(t)$. We have
\begin{align*}
\norm{J(t)}_{L^4} &\le C \int_{t_1}^t |t-s|^{-3/4}
\norm{G_\xi(s)}_{L^{4/3}} \, d s \le  C \int_{t_1}^t |t-s|^{-3/4}
n^{-\iota} \nxi {s}^{-3/2} \, d s
\\
&\le C n^{-\iota} \nxi t_1^{-1/2} t^{-3/4} \le C \nxi^{2/3}
n^{(2+\delta)/3 - \iota}  t^{-3/4}\ll \nxi t^{-3/4},
\end{align*}
where we have used the inequality \eqref{cal-3} to bound the last
integral, and also $\nxi^{1/3} t_1^{-1/2} = n^{(2+\delta)/3}$ by
\eqref{3-15}. Similarly,
\begin{align*}
\norm{J(t)}_{L^2\loc}& \le C\int_{t_1}^{t} \min \bket{
|t-s|^{-3/2}, \ |t-s|^{-3/4} } \norm{G_\xi(s)}_{L^1 \cap L^{4/3}}
\, d s
\\
&\le C\int_{t_1}^{t} \min \bket{ |t-s|^{-3/2}, \ |t-s|^{-3/4} }
n^{-\iota} \nxi {s}^{-3/2} \, d s \le C n^{-\iota} \nxi {t}^{-3/2}
\end{align*}
by \eqref{cal-4}. We have proven the estimates of $\xi(t)$ in
\eqref{t2-est}.

We will estimate $x$ and $y$ using the normal form in Lemma
\ref{th:NF} with the initial time $t=t_1$. Recall that
$x(t)=e^{-ie_0t}u(t)$, $y(t)=e^{-ie_0t}v(t)$ and the perturbations
$\mu$ of $u$ and $\nu$ of $v$ satisfy \eqref{|mu|} and
\eqref{|nu|}. We first estimate the error terms $g_u$ and $\wt
g_v$ in \eqref{|mu|}--\eqref{|nu|}, for which we need a bound on
$\norm{\xi^{(3)}(t)}_{L^2\loc} $.

Recall $\xi^{(3)}=\xi^{(3)}_1+\xi^{(3)}_2+\xi^{(3)}_{3-5} $ is
defined in \eqref{xi3.def}. We set the initial time to $t_1$ and
replace $\xi_0$ by $\xi(t_1)$ in \eqref{xi3.def}. The estimate of
$\xi^{(3)}_1$ is given by \eqref{out:t1}.
$\norm{\xi^{(3)}_2(t)}_{L^2\loc}$ is bounded by $C n^2|x(t_1)|
\bka{t-t_1}^{-3/2}$ by Lemma \ref{th:2-2} and the definition
\eqref{2-17} of $\xi^{(2)}(t_1)$. For $\xi^{(3)}_{3-5}$ we have
the integral estimate \eqref{xi335.est} and we can use Lemma
\ref{th:Gxi} to bound the integrand,
 $g_{\xi,3-5}(s) \lesssim n^4 x + X \le C(n^{2-\iota}\nxi+
n^{1-\iota}\nxi^2 + \nxi^3) s^{-3/2} = o(1) \nxi s^{-3/2}$.
Summing all the estimates, we can bound $\xi^{(3)}(t)$ by
\begin{align*}
\norm{\xi^{(3)}(t)}_{L^2\loc} &\le (1+\iota) \nxi t^{-3/2} + C
n^2n^{2+\delta} \bka{t-t_1}^{-3/2}
\\
&\quad + \int_{t_1}^t \min \bket{|t-s|^{-3/2},|t-s|^{-3/4}} o(1)
\nxi s^{-3/2} d s
\\
&\le (1+\iota) \nxi t^{-3/2} + C n^{2} t^{-3/2} + o(1) \nxi
t^{-3/2} \le 2 \nxi t^{-3/2}.
\end{align*}
Here we have used $n^{2+\delta}\bka{t-t_1}^{-3/2} \le C t^{-3/2}$
for $t\ge t_1$. Using \eqref{guv.est}, \eqref{3-22}, \eqref{3-14}
and \eqref{3-5}, we can bound the error terms $g_u$ and $\wt g_v$
by
\begin{align*}
|g_u|,|\wt g_v| &\lesssim \nxi n^5 |x| + n^2 \norm{\xi^{(3)}}_{L^2\loc}
+ n \norm{\xi}_{L^2\loc}^2  +
  ( \norm{\xi}_{L^2\loc}+ \nxi n^2) \norm{\xi}_{L^4}^2
\\
&\lesssim n^2 \nxi t^{-3/2}.
\end{align*}

We now estimate $x(t)$. If $t>n^{-10/3}$, using $\delta + \iota <
1$ we have $|x(t)|\le n^{-2-\iota} \nxi t^{-3/2} \le n^{-2-\iota}
\nxi (n^{-10/3})^{-3/2} \le n^{2+\delta}$.

If $t \in [t_1,n^{-10/3}]$, using \eqref{|mu|} we have,
\begin{align}
\big| |\mu(t)|-|\mu(t_1)| \big|  &\le  \int_{t_1}^{t} \abs{ \frac
d{ds}|\mu(s)| } \, d s  \nonumber
 \le C \int_{t_1}^{t}  n^4 |x(s)| +  |g_u(s)| \, d s
\\
&\le C\int_{t_1}^{t} n^4 n^{2+\delta} + n^2 \nxi s^{-3/2}
\, d s \nonumber
\\
&\le C n^{6+\delta} n^{-10/3} + C n^{2} \nxi \bka{t_1}^{-1/2} \ll
n^{2+\delta}. \label{3-25}
\end{align}
Here we have used $\nxi^{1/3} t_1^{-1/2} = n^{(2+\delta)/3}$ and
$\delta<1$. Since $\mu = u + O(n^2 u)$, together with \eqref{3-22}
we have proved the estimate for $x(t)$ in \eqref{t2-est}.

We now estimate $y(t)$. Using \eqref{|nu|} and \eqref{3-22}, for
all $t \in [t_1,t_2)$ we have
\begin{align}
\big| |\nu(t)|-|\nu(t_1)| \big|  &\le \int_{t_1}^{t} \abs{ \frac
d{ds}|\nu(s)| } \, ds
 \le C \int_{t_1}^{t}  n^4 |x(s)| +  |\wt g_\nu(s)| \, d s
 \nonumber
\\
&\le C \int_{t_1}^{t} n^{2-\iota} \nxi s^{-3/2}  \, d s 
\le C n^{2-\iota} \nxi \bka{t_1}^{-1/2} \ll n. \label{3-26}
\end{align}
Since $\nu = v + O(n^2 u)$, we have proved the estimate for $y(t)$
in \eqref{t2-est}. The proof of \eqref{t2-est} is complete.

Suppose $t_2=\infty$. The bounds of $x(t)$ and
$\norm{\xi(t)}_{L^2\loc}$ in \eqref{3-18} are given by
\eqref{t2-est}. By the same argument as in \eqref{3-26}, we have
for all $t>\tau \ge t_1$,
\[
\big| |\nu(t)|-|\nu(\tau)| \big|  \le C \int_{\tau}^{t}
n^{2-\iota} \nxi s^{-3/2} \, d s \le C n^{2-\iota} \nxi
\bka{\tau}^{-1/2} ,
\]
which converges to zero as $t,\tau \to \infty$. Hence $|\nu(t)|$
and $|y(t)|$ have a limit $y_\infty$.  Moreover, $\big|
|y_\infty|-|\nu(\tau)| \big| \le C \tau^{-1/2}$  as $\tau \to
\infty$ and $y_\infty-|y(t_1)|$ is bounded by $C n^{2-\iota} \nxi
\bka{t_1}^{-1/2} \ll n$. Hence $y_\infty \sim n$. The phase
$\Theta(t)$ of $y(t)$ is given in \eqref{y.def}, $ \Theta(t)=
\theta(t) - \int_0^t E_1(|y(s)|)ds$. Let $E_\infty=E_1(y_\infty)$.
Using \eqref{mt.eq} we have
\begin{align*}
\abs{\Theta(t) + E_\infty t} &\le |\theta(0)|+\int_0^t |\dot
\theta| + \abs{E_1(y(s)) - E_\infty} d s \\ &\le C + \int_0^t C
n^{-1} \norm{G(s)}_{L^1\loc} + C n^2 \abs{|y(s)| - y_\infty} d s
\\
&\le C + \int_0^t C (1+s)^{-1/2} d s \le C (1+t)^{1/2}.
\end{align*}
We have completed the proof of \eqref{3-18}.

Suppose now  $t_2 < \infty$. For $t\ge t_2$ we have
\[
e^{-i(t-t_2) H_0}\xi(t_2) = e^{-i (t-t_1) H_0} \xi(t_1) +  J_2(t),
\]
where
\[
 J_2(t) =  \int_{t_1}^{t_2} e^{-i (t- s ) H_0}
\PcH G_\xi( s ) \, d  s .
\]
The estimate for $e^{-i (t-t_1) H_0} \xi(t_1)$ is provided by
\eqref{out:t1} of Proposition \ref{th:t1}. Hence we only need to
estimate $ J_2(t)$. Recall $\norm{ G_\xi(s)}_{L^{4/3}\cap L^1}\le
C n^{-\iota} \nxi \bka{ s }^{-3/2}$ for $ s  \in [t_1,t_2]$.
Hence, by \eqref{cal-3} and \eqref{3-15},
\begin{align*}
\norm{ J_2(t) }_{L^4} &\le C\int_{t_1}^{t_2} |t - s|^{-3/4} \norm{
G_\xi( s )}_{L^{4/3}} \, d  s  \le C \int_{t_1}^t |t- s |^{-3/4}
n^{-\iota} \nxi s ^{-3/2} d  s
 \\
 &\le C
n^{-\iota} \nxi t_1^{-1/2} \bka{t }^{-3/4} \le C \nxi^{2/3}
n^{(2+\delta)/3 - \iota} t^{-3/4}\ll \nxi t^{-3/4}.
\end{align*}
This proves the first bound in \eqref{out:t2}.

For the $L^2\loc$ norm we have
\begin{align*}
\norm{J_2(t)}_{L^2\loc} &\le \int_{t_1}^{t_2} \Omega(s) \,  d s
\le \int_{\ell} ^{t-\ell} \Omega(s) ds +\int_{t_2- \ell}^{t_2} \Omega(s) ds,
\end{align*}
where  $\ell = t_1/2$ and
$\Omega(s): = \min \bket{|t-s|^{-3/2}, |t-s|^{-3/4}} C n^{- \iota} \nxi s^{-3/2} $.
We have
\[
\int_{\ell} ^{t-\ell} \Omega(s) ds
\le C n^{- \iota} \nxi  \int_{\ell}^{t-\ell} |t-s|^{-3/2}  s^{-3/2} ds
= 2 C n^{- \iota} \nxi  \int_\ell^{t/2} |t-s|^{-3/2}  s^{-3/2} ds.
\]
The last integral is bounded by
\[
\int_\ell ^{t/2} t^{-3/2} C n^{- \iota} \nxi s^{-3/2} ds \le C
n^{- \iota} \nxi \ell ^{-1/2} t^{-3/2}   \ll \nxi t^{-3/2}.
\]
Recall  $\nxi t_2^{-3/2} = n^{2+\iota}|x(t_2)|$. Since
$s\sim t_2$ for $s\in
[t_2 -\ell,t_2]$, by
\eqref{cal-2} we can bound
the second integral by
\begin{align*}
& \int_{t_2- \ell}^{t_2} \Omega(s) ds \le \int_{t_2-\ell} ^{t_2} \min \bket{|t-s|^{-3/2},
|t-s|^{-3/4}} C n^{- \iota} \nxi t_2^{-3/2} d s
\\
&\le C n^{2}|x(t_2)| \frac {\ell}{\ell + t-t_2}
\bka{t-t_2}^{-1/2}.
\end{align*}
Combining these two bounds, we have proved
the second bound in \eqref{out:t2}.
\myendproof

The significance of $t_2$ is that it is a time when the dispersion
loses its dominance over the ground state.
\donothing{
In terms of estimate, the right side of \eqref{out:t2}$_2$ for the
$L^2\loc$-norm consists of two parts, the first once generated from
$t\sim 0$ and the second from $t\sim t_2$. The first part decays
slowly for $t>t_2$ but is already smaller than $n^2 x(t_2)$, the
size of $\xi^{(2)}$. Hence it is negligible for all $t\ge t_2$.
The second part has the same size as $n^2 x(t_2)$ but decays
quickly as $t$ increases, and hence is negligible for $t>t_2 +
\delta t$.
}
If $t_2=\infty$, by \eqref{3-18} the solution $\psi(t)$
converges locally to an excited state $Q_1(y_\infty)$. This gives us case
III of Theorem \ref{th:1-2}.  We shall consider the other case
$t_2 < \infty$, which corresponds to case II$_b$, in Propositions
\ref{th:t3}--\ref{th:t4}.

\begin{proposition}  \label{th:t3} Suppose that the assumptions
of Proposition \ref{th:t2} hold and $t_2<\infty$. Let
\[
t_3\equiv \inf \bket{t\ge t_2:\ |x(s)| < 0.001n ,\ \forall s \in
[t_2,t]}.
\]
We have
 \[
 t_2 + n^{-4} \le t_3 < \infty, \qquad |x(t_3)|=0.001n,
 \]
and the following estimates for all $s,t$ with $t_2\le s\le t \le
t_3$:
\begin{equation} \label{t3:est}
\begin{split}
|y(t)/y(t_2)| &\in  [\tfrac {24}{25},\ \tfrac {26}{25}],
\\
|x(t)/x(s)| &\in [\tfrac 34 \ e^{\tfrac 34  \gamma_0 n^4(t-s)} ,\
\tfrac 54 \ e^{\tfrac 54 \gamma_0 n^4 (t-s)}],
\\
\norm{\xi(t)}_{L^4} &\le  C_3\, n|x(t)| + (1+3\iota) \nxi t^{-3/4}
,
\\
\norm{\xi(t)}_{L^2 \loc} &\le  C_3\, n^2 |x(t)| + (1+3\iota) \nxi
t^{-3/2} ,
\\
\norm{\xi^{(3)}(t)}_{L^2\loc} &
\le C_3 \nxi n^3 |x(t)| + (1+3\iota) \nxi t^{-3/2} \\
&\quad + C_3  n^{2}|x(t_2)| \frac {t_1}{t_1+ t-t_2}
\bka{t-t_2}^{-1/2} ,
\end{split}
\end{equation}
for some explicit constant $C_3>0$.

\end{proposition}

\myproof  Using a continuity argument we may assume
\begin{equation} \label{3-23}
\begin{split}
|y(t)/y(t_2)| &\in [\tfrac {19}{20},\ \tfrac {21}{20}]
\\
|x(t)/x(s)| &\in [\tfrac 12 \ e^{\tfrac 12  \gamma_0 n^4(t-s)} ,\
\tfrac 32 \ e^{\tfrac 32 \gamma_0  n^4 (t-s)}]
\\
\norm{\xi(t)}_{L^4} &\le 2 C_3\, n|x(t)| + 2 \nxi t^{-3/4} ,
\\
\norm{\xi(t)}_{L^2 \loc} &\le  2C_3\, n^2 |x(t)| + 2 \nxi
t^{-3/2},
\\
\norm{\xi^{(3)}(t)}_{L^2\loc} & \le 2C_3 \nxi n^3 |x(t)| + 2 \nxi
t^{-3/2} +
 \frac {2C_3  n^{2}|x(t_2)|t_1}{t_1+ t-t_2}
\bka{t-t_2}^{-1/2} .
\end{split}
\end{equation}
We now apply Lemma \ref{th:NF} with {\it the initial
time set to $t=t_2$} to obtain the normal form for $x$ and $y$
and the decomposition of $\xi$.

We first estimate $G_\xi$ and $g_{\xi,3-5}$. By Lemma
\ref{th:Gxi}, $\norm{G_\xi(s)}_{L^1\cap L^{4/3}} \lesssim n^2 x +
X$ with
\[
X(s) = n \nxi \norm{\xi}_{L^2\loc} + \nxi \norm{\xi}_{L^4}^2 \le
C\nxi n^3 |x(s)| + C\nxi^3 s^{-3/2}.
\]
Hence $\norm{G_\xi(s)}_{L^1\cap L^{4/3}} \le Cn^2|x(s)| + \nxi^3
s^{-3/2}$. By the same Lemma $g_{\xi,3-5}(s) \lesssim n^4 |x(s)| +
X \le C\nxi n^3|x(s)| + \nxi^3 s^{-3/2}$. Note we have $|x(s)|\le
2|x(t)|e^{-\tfrac 12  \gamma_0 n^4(t-s)} $ for $t_2\le s \le t \le
t_3$ by \eqref{3-23}.

We now estimate $\xi(t)$. For $t\in [t_2,t_3]$ we have
\[
\xi(t) = e^{-i (t-t_2) H_0} \xi(t_2) + J(t), \qquad J(t)\equiv
\int_{t_2}^t  e^{-i (t-s) H_0} \PcH G_\xi(s) \, d s .
\]
The estimate for $e^{-i (t-t_2) H_0} \xi(t_2)$ is by
\eqref{out:t2} of Proposition \ref{th:t2}. Hence it suffices to
estimate the integral. By the above estimate of $G_\xi$ we have,
using \eqref{cal-3},
\begin{align*}
\norm{J(t)}_{L^{4}}& \le C \int_{t_2}^t |t-s|^{-3/4}
\norm{G_\xi(s)}_{L^{4/3}} \, d s
\\
&\le C \int_{t_2}^t |t-s|^{-3/4} \bke{ n^2|x(t)|e^{-\tfrac 12
\gamma_0 n^4(t-s)}  + \nxi^3 s^{-3/2} }\, d s
\\
& \le C n^2 |x(t)|(n^{-4})^{1/4} + C \nxi^3 t_2^{-1/2} {t}^{-3/4}
\le C n |x(t)| + o(1) \, \nxi {t}^{-3/4}.
\end{align*}
For the $L^2\loc$ norm, since $\xi =\xi^{(2)} +\xi^{(3)}$ and
$\norm{\xi^{(2)}(t)}_{L^2\loc} \le C n^2 |x(t)|$ by its explicit
form, it suffices to estimate $\xi^{(3)} =\xi^{(3)}_1+
\xi^{(3)}_2+ \xi^{(3)}_{3-5}$. By the above estimate of
$g_{\xi,3-5}$ and \eqref{xi335.est},
\begin{align*}
\norm{\xi^{(3)}_{3-5}(t)}_{L^2\loc} &\le \int_{t_2}^t \min
\bket{|t-s|^{-3/2},|t-s|^{-3/4}} g_{\xi,3-5}(s)\, d s
\\
& \le C \int_{t_2}^t \min \bket{|t-s|^{-3/2},|t-s|^{-3/4}}
\\
& \qquad \qquad \cdot
\bke{\nxi n^3|x(t)|e^{-\tfrac 12  \gamma_0 n^4(t-s)} + \nxi^3
s^{-3/2}}\, d s
\\
&\le C \nxi n^3|x(t)| + C \nxi^3 t^{-3/2}.
\end{align*}
The estimate of $\xi^{(3)}_1(t)$ is given in \eqref{out:t2}. We
also have
\[
\norm{\xi^{(3)}_2(t)}_{L^2\loc} \le C n^2 |x(t_2)|
\bka{t-t_2}^{-3/2} \le C n^2 |x(t_2)| \frac {t_1}{t_1+ t-t_2}
\bka{t-t_2}^{-1/2}
\]
by its explicit form and Lemma \ref{th:2-2}. Hence the
$L^2\loc$-bounds of $\xi$ and $\xi^{(3)}$ are proved.

We next estimate $g_u$ and $\wt g_v$. By \eqref{guv.est} of Lemma
\ref{th:NF}, \eqref{3-23}, and $\nxi t^{-3/2} \le n^{2+\delta}$,
\begin{align*}
|g_u|,|\wt g_v| &\lesssim \nxi n^5 |x| + n^2 \norm{\xi^{(3)}}_{L^2\loc}
+ n \norm{\xi}_{L^2\loc}^2  +
  ( \norm{\xi}_{L^2\loc}+ \nxi n^2) \norm{\xi}_{L^4}^2
\\
&\lesssim  \nxi n^5 |x(t)| +  \nxi n^2 t^{-3/2} +
  n^{4}|x(t_2)| \frac {t_1}{t_1+ t-t_2} \bka{t-t_2}^{-1/2}.
\end{align*}

We now estimate $x(t)$ and $y(t)$ using the normal form
\eqref{eq:NF} in Lemma \ref{th:NF}  for the perturbation $\mu(t)$
of $u(t)=e^{i e_0 t} x(t)$ and $\nu(t)$ of $v(t)=e^{i e_1 t}
y(t)$. Recall that the initial time for normal form and the
decomposition of $\xi$ is reset at $t=t_2$.
We first consider $t\in [t_2,t_2+n^{-1}]$. Using the estimate of
$g_u$ and $\al t_2^{-3/2} = n^{2+\iota}|x(t_2)|$,
\[
\abs{\frac d{dt}|\mu|} \lesssim n^4 |x| + |g_u| \lesssim n^4 |x| +
\al n^2 t^{-3/2} \lesssim   n^4 |x(t_2)| .
\]
Therefore $\big||\mu(t)| - |\mu(t_2)|\big|\le C n^4 |x(t_2)|
(t-t_2)\ll |x(t_2)|$, and hence $|x(t)|\sim |x(t_2)|$. Similarly
$|y(t)|\sim |y(t_2)|$.

We next consider $t \in [t_2+n^{-1},t_3]$. The previous estimate
of  $g_u$ and $\wt g_v$ becomes
\begin{equation}\label{3-3:guv.est}
|g_u|,|\wt g_v| \lesssim  \nxi n^5 |x(t)| +  \nxi n^2 t^{-3/2} +
n^{4+1/2}|x(t_2)|\le C n^{4+\iota} |x(t)|.
\end{equation}
Here we have used $\nxi t^{-3/2}\le n^{2+\iota}|x(t_2)|$ and
$|x(t_2)|\le |x(t)|$.

By the estimate of $g_u$ and \eqref{|mu|}, we have
\[
|\mu| ^{-1} \frac d{ dt} |\mu| \in [ \tfrac 34  \gamma_0 n^4 , \
\tfrac 54 \gamma_0 n^4 ] .
\]
Hence, for all $s, t$ with $t_2\le s \le t \le t_3$,
\[
|\mu(t)/\mu(s)| \in [ e^{\tfrac 34  \gamma_0 n^4(t-s)} ,\
e^{\tfrac 54 \gamma_0  n^4 (t-s)}].
\]
Since $\big| |x|-|\mu| \big| \le C n^2 |x|$, we have proven
\eqref{t3:est}$_2$ for $|x(t)/x(s)|$.
Since $|x(t)|\le 0.001 n$ for all $t<t_3$, we must have $t_3 <
\infty$. Moreover,
\begin{equation}
(\tfrac 54 \gamma_0  n^4)^{-1} \ \log \frac {4|x(t_3)|}
{5|x(t_2)|} \le t_3 - t_2 \le (\tfrac 34 \gamma_0  n^4)^{-1} \
\log \frac {4|x(t_3)|} {3|x(t_2)|}.
\end{equation}

We now estimate $|y(t)/y(s)|$. By the estimate of $\wt g_v$ and
\eqref{|nu|}, we have
\begin{equation} \label{3-40}
\left| \frac d{ dt} |\nu| \right |
 = \left|  -2 \gamma_0  |\mu|^2|\nu|^3 + \Re \wt g_v \bar \nu/|\nu|
 \right | \le 3 \gamma_0 n^4 |x(s)|.
\end{equation}
By \eqref{3-23}
\begin{align*}
\big| |\nu(t)|-|\nu(t_2)| \big| &\le \int_{t_2}^t  3 \gamma_0 n^4
|x(s)|\, d s \le \int_{t_2}^t  3 \gamma_0 n^4 2|x(t)|e^{-\tfrac 12
\gamma_0 n^4 (t-s)} \, d s
\\
&\le 12 |x(t)| \le 0.012 n.
\end{align*}
Since $\big| |y|-|\nu| \big| \le C n^2 |x|$, we have proven
\eqref{t3:est}$_1$ for $|y(t)/y(s)|$.
\myendproof

\begin{proposition}  \label{th:t4} Assume the same assumptions of
Proposition \ref{th:t2}. Let
\begin{equation}
t_4\equiv \sup \bket{t\ge t_3:\ |y(s)| \ge \e n, \quad \forall s
\in [t_3,t)},
\end{equation}
where $\e=\e_0/4$ and $\e_0>0$ is the small constant in Theorem
\ref{th:1-1}. We have
\begin{equation} \label{eq:3-4-1}
t_3 \le t_4 \le  t_3 + C  (\gamma_0 \e^2 n^4)^{-1}.
\end{equation}
We also have the following estimates for $t_3\le t\le t_4$:
\begin{equation} \label{eq:3-8B}
\begin{split}
&\tfrac 1{1200} n \le |x(t)|\le 2 n, \qquad |y(t)| \le 2 n,\\
& |x(t_4)| \ge \tfrac 12 n  , \qquad |y(t_4)| = \e n .
\end{split}
\end{equation}
\begin{equation}\label{xi.est}
\begin{split}
\norm{\xi(t)}_{L^4} &\le C_4\, n|x(t)| + (1+4\iota) \nxi t^{-3/4}
,
\\
\norm{\xi(t)}_{L^2 \loc} &\le  \, C_4\,  n^{2} |x(t)| + (1+4\iota)
\nxi t^{-3/2} ,
\\
\norm{\xi^{(3)}(t)}_{L^2 \loc} &\le C_4 n^{2+\iota} |x(t)|,
\end{split}
\end{equation}
for some explicit constant $C_4>0$.
Moreover, for $t\ge t_4$ and $\dt = \e^{-2} n^{-4}$,
we have the following outgoing estimates
on the dispersive wave $\xi$:
\begin{equation}\label{out:t4}
\begin{split}
\norm{ e^{-i(t-t_4)H_0} \xi(t_4)}_{L^4} & \le C  n^3 \dt (\dt
+t-t_4)^{-3/4} ,
\\
\norm{ e^{-i(t-t_4)H_0} \xi(t_4)}_{L^2 \loc} & \le  C  n^3
\frac{\dt}{\dt + t-t_4} \ (1+t-t_4)^{-1/2} .
\end{split}
\end{equation}
Hence the conditions of Theorem \ref{th:1-1} are satisfied at
$t=t_4$ and the solution $\psi(t)$ converges locally to a
nonlinear ground state.

\end{proposition}

\myproof  By a continuity argument and \eqref{eq:3-2}, we may
assume
\begin{equation}\label{t4-asp}
\begin{split}
\tfrac 1{1400} n &\le |x(t)|\le 2 n, \qquad |y(t)| \le 2 n,
\\
\norm{\xi(t)}_{L^4} &\le 2C_4\, n|x(t)| + 2 \nxi t^{-3/4} ,
\\
\norm{\xi(t)}_{L^2 \loc} &\le  \,2 C_4\,  n^{2} |x(t)| + 2 \nxi
t^{-3/2},
\\
\norm{\xi^{(3)}(t)}_{L^2 \loc} &\le 2 C_4 n^{2+\iota} |x(t)| .
\end{split}
\end{equation}

The estimates for $G_\xi$, $\xi$, $g_{\xi,3-5}$, $\xi^{(3)}$,
$g_u$ and $\wt g_v$ can be proved in the same way as those in
Proposition \ref{th:t3}. The only difference is on the estimates
of the bound states $x$ and $y$, which we now focus on.

For any $t\le t_4$, we have $|y(t)|\ge \e n$. By \eqref{|mu|},
\eqref{3-3:guv.est} and \eqref{t4-asp}, we have
\[
\frac d{ dt} |\mu| = \gamma_0 |\nu|^4 |\mu|+ \Re g_\mu \bar \mu
/|\mu| \ge \gamma_0 \e^4 n^4 |\mu| - C n^{4+\iota} |\mu| \ge
\tfrac 78 \gamma_0 \e^4 n^4 |\mu|.
\]
Hence $|x(t)|, |\mu(t)| \ge |x(t_3)| (1 - C n^2) \ge \tfrac
1{1200}n$. By \eqref{|nu|}, \eqref{3-3:guv.est}, \eqref{t4-asp},
and $\nu = v + O(n^2x)$,
\begin{align*}
\frac d{ dt} |\nu| &= -2 \gamma_0  |\mu|^2|\nu|^3 + \Re \wt g_v
\bar \nu/|\nu|
\\
&\le - 2 \gamma_0 (\tfrac n{1400})^2 |\nu|^3 + C n^{5+\iota} < -
10^{-6}  \gamma_0 n^2 |\nu|^3.
\end{align*}
Let $\rho(t) =\bket{ |\nu(t_3)|^{-2} + 2 (10^{-6}) \gamma_0 n^2 (t
-t_3) } ^{-1/2}$. We have $\rho(t_3)=|\nu(t_3)|$ and $\frac d{ dt}
\rho =- 10^{-6}  \gamma_0 n^2 \rho^3$. By comparison principle,
\[
|\nu(t)| \le \rho(t)=\bket{ |\nu(t_3)|^{-2} + 2 (10^{-6}) \gamma_0
n^2 (t -t_3) } ^{-1/2}.
\]
Since $|\nu(t)|\ge \e n/2$ for $t\in [t_3,t_4]$, we have $t_4 <
\infty$ and  $t_4-t_3 \le C \gamma_0^{-1}\e^{-2} n^{-4}$.
Similarly,
\[
\frac d{ dt} |\nu| \ge -3 \gamma_0 n^2 |\nu|^3.
\]
We can compare it with $\rho_-(t)=\bket{ |\nu(t_3)|^{-2} + 6
\gamma_0 n^2 (t -t_3) } ^{-1/2}$ and obtain $\nu(t) \ge \rho_-(t)$
by comparison principle. This gives a lower bound $t_4-t_3\ge C
\dt$ if $|\nu(t_3)| \ge 2 \e n$.

We finally prove the outgoing estimates \eqref{out:t4} for
$\xi(t_4)$. For $t \ge t_4$,
\begin{equation*} 
e^{-i (t-t_4) H_0}\xi(t_4) = e^{-i (t-t_2) H_0}\xi(t_2) + J_4(t),
\end{equation*}
where $ J_4(t)$ denotes the integral
\begin{equation*} 
 J_4(t) = \int_{t_2}^{t_4} e^{-i(t -s)H_0} \PcH
G_\xi(s) \, d s  .
\end{equation*}
The estimate of $e^{-i (t-t_2) H_0}\xi(t_2)$ is by \eqref{out:t2}
of Proposition \ref{th:t2}. Hence we only need to estimate $
J_4(t)$. \donothing{ By \eqref{3-49}, $ J_4(t)$ is the difference
of two $L^2$ functions and hence $\norm{ J_4(t)}_{L^2} \le
\norm{\xi(t_4)}_{L^2} + \norm{\xi(t_2)}_{L^2} \le 2 \nxi$ for all
$t$.}
%
We have shown in the proof of Proposition \ref{th:t3} that, for $s
\in [t_2,t_3]$,
\[
\norm{G_\xi(s)}_{L^1\cap L^{4/3}} \le n^2|x(t_3)|e^{-\tfrac 12
\gamma_0 n^4(t_3-s)}  + \nxi^3 s^{-3/2}, \qquad (t_2\le s \le
t_3).
\]
For $s \in [t_3,t_4]$, using $t_3 \ge n^{-4}$,
\[
\norm{G_\xi(s)}_{L^1\cap L^{4/3}} \le C n^3,\qquad (t_3\le s \le
t_4).
\]
By the exponential decay, \eqref{cal-4}, \eqref{cal-3} and
$t_4-t_3 \le C \dt$, $\dt= \e^{-2} n^{-4}$,
\begin{align*}
\norm{  J_4(t) }_{L^4} &\le C \int _{t_2}^{t_4} |t-s|^{-3/4}
\norm{G_\xi(s)}_{L^{4/3}} \, d s
\\
&\le \int _{t_2}^{t_3} |t-s|^{-3/4} \bke{n^2|x(t_3)|e^{-\tfrac 12
\gamma_0 n^4(t_3-s)}+ \nxi^3 s^{-3/2}} d s
 \\
&\qquad + \int _{t_3}^{t_4} |t-s|^{-3/4} Cn^3 ds
\\
&\le C \int _{t_3 - n^{-4} } ^{t_3} |t-s|^{-3/4} n^3 \, d s +   C
\nxi^3 t_2^{-1/2} t^{-3/4} +C \int _{t_3 } ^{t_4} |t-s|^{-3/4} n^3
\, d s
\\
&\le C n^3 \dt (\dt + t-t_4)^{-3/4} + C \nxi^3 t_2^{-1/2}
t^{-3/4}.
\end{align*}
Since $\nxi t^{-3/4} \le C n^3 \dt (\dt + t-t_4)^{-3/4}$ for $t\ge
t_4$, we have the $L^4$ estimate in \eqref{out:t4}.
For $L^2\loc$-norm, we have
\begin{align*}
\norm{  J_4(t) }_{L^2 \loc}  &\le C \int _{t_2}^{t_4} \min \bket{
|t- s |^{-3/2} , \, |t- s |^{-3/4} } \norm{G_\xi(s)}_{L^1\cap
L^{4/3}} \, d  s
\\
&\le \int _{t_2}^{t_3} |t-s|^{-3/2} \bke{n^2|x(t_3)|e^{-\tfrac 12
\gamma_0 n^4(t_3-s)}+ \nxi^3 s^{-3/2}} d s
\\
&\qquad + \int _{t_3}^{t_4} \min \bket{ |t- s |^{-3/2} , \, |t- s
|^{-3/4} } Cn^3 d s
\\
&\le C \int _{t_3 - n^{-4} } ^{t_3} |t-s|^{-3/2} n^3 \, d s
+   C \nxi^3 t^{-3/2}
\\
& \qquad  + C n^3 \frac {\dt}{\dt + t -t_4} (1+t-t_4)^{-1/2}
\\
&\le C n^3 \frac {\dt}{\dt + t -t_4} (1+t-t_4)^{-1/2}.
\end{align*}
Here we have used the exponential decay, \eqref{cal-2},
\eqref{cal-1}, and that  $\nxi^3 t^{-3/2}$ is less than the last
quantity for $t\ge t_4\ge n^{-4}$. The proof is complete.
\myendproof

We now come back to case II$_a$ where  $t=t_1$ and we have $|x(t_1)|
\ge |y(t_1)|^{2+\delta}$. We further divide it to three subcases:

1. $|y(t_1)|=n$, $n^{2+\delta}\le |x(t_1)|\le 0.001n$.

2. $\max(|x(t_1)|, |y(t_1)|)=n$, $ 0.001n\le |x(t_1)| \le n$,
$\e_0 n\le |y(t_1)|\le n$.

3. $|x(t_1)|=n$, $|y(t_1)|\le \e_0 n$.
\\
In case 1, we can set $t_2=t_1$ and our analysis in Propositions
\ref{th:t3}--\ref{th:t4} and Theorem \ref{th:1-1} for $t\in
[t_2,\infty)$ goes through. In case 2, we set $t_3=t_2=t_1$ and
apply Proposition \ref{th:t4} and Theorem \ref{th:1-1}. In case 3,
we can set $t_4=t_1$ and apply Theorem \ref{th:1-1} directly.
The proof of Theorem \ref{th:1-2} is complete.

\subsection*{Acknowledgments}
Tsai was partially supported by NSF grant DMS-9729992.
Yau was partially supported by NSF grant DMS-0072098.

\end{document}